\documentclass[preprint,preprintnumbers,showpacs,aps,prd,amssymb]{revtex4-1}

\usepackage{graphicx}
\usepackage{bm}
\usepackage{amsmath}

\def\calC{{\cal C}}
\def\calD{{\cal D}}
\def\calH{{\cal H}}
\def\calO{{\cal O}}
\def\calS{{\cal S}}

\def\Bbar{{\bar B}}

\def\cbar{{\bar c}}
\def\hbar{{\bar h}}

\def\rhat{{\hat r}}

\def\Ds{D^{(*)}}

\def\nn{\nonumber}

\begin{document}
%
\title{$B$ anomalies in the nonminimal universal extra dimension model}
\author{Jong-Phil Lee}
\email{jongphil7@gmail.com}
\affiliation{Sang-Huh College,
Konkuk University, Seoul 05029, Korea}

\begin{abstract}
We investigate the $B$ anomalies in the framework of the nonminimal universal extra dimension models.
Newly measured polarization parameters in $B\to\Ds\tau\nu$,  $P_\tau(\Ds)$ and $F_L(D^*)$
as well as the ratios $R(\Ds)$ are considered altogether.
The Kaluza-Klein modes of the $W$-boson and charged scalar contributes as the new physics effects.
We find that the model parameters fit the global data very well with the minimum $\chi^2/{\rm d.o.f.}$ 
near unity, rendering $B_c\to\tau\nu$ branching ratios to be a few percents.
The best-fit values of $R(D)$ and $R(D^*)$ are still far from ($\gtrsim 2\sigma$) the standard model predictions.
\end{abstract}

\maketitle

\section{Introduction}
The standard model (SM) of particle physics has been up to now very successful to explain 
many phenomena in our universe.
The last missing piece of the SM, the Higgs particle was finally discovered in 2012.
But there must be some new physics (NP) beyond the SM.
Flavor physics is a good testing ground for the NP.
Recently, some anomalies are reported in $b\to c$ semileptonic decays.
The fraction of the branching ratios
\begin{equation}
R(\Ds)\equiv\frac{{\rm Br}(B\to \Ds\tau\nu)}{{\rm Br}(B\to \Ds\ell\nu)}~,
\end{equation}
reveals an excess over the SM predictions \cite{Amhis},
\begin{eqnarray}
R(D)_{\rm SM} &=&0.299\pm0.003~,\nonumber\\
R(D^*)_{\rm SM} &=&0.258\pm0.005~.
\label{RDSM}
\end{eqnarray}
Experiments including BABAR, Belle, and LHCb have reported somewhat larger values of $R(\Ds)$
than those of Eq.\ (\ref{RDSM}) by about $2\sim3\sigma$ 
\cite{BaBar_PRL,BaBar1,Belle1,Belle1607,Belle1703,Belle1612,Belle1709,Belle1904,LHCb1,LHCb2}.
Recently the Belle collaboration announced new results \cite{Belle1904}
\begin{eqnarray}
R(D)_{\rm Belle1904} &=&0.307\pm0.037\pm0.016~,\nonumber\\
R(D^*)_{\rm Belle1904} &=&0.283\pm0.018\pm0.014~,
\end{eqnarray}
which are rather closer to Eq.\ (\ref{RDSM}) than the previous data and consistent with the SM
within $1.2\sigma$.
Combined results for all data by the heavy flavor averaging group (HFLAV) collaboration \cite{HFAG2019}
\begin{eqnarray}
R(D)_{\rm HFLAV} &=& 0.340\pm0.027\pm0.013~,\nn\\
R(D^*)_{\rm HFLAV} &=& 0.295\pm0.011\pm0.008~,
\label{RDHFLAV}
\end{eqnarray}
give a discrepancy between the SM predictions and experimental data at $3.08\sigma$ level.
The BABAR measurements \cite{BaBar_PRL,BaBar1} exclude at the 99.8\% confidence level
the type-II two-Higgs-doublet model (2HDM) where a charged Higgs boson contributes to $R(\Ds)$, 
while the Belle measurements \cite{Belle1} are compatible with the type-II 2HDM.
It was shown that an anomalous $\tau$ coupling to the charged Higgs in the 2HDM can explain 
the data very well \cite{jplee}.
In extra dimension models the overlapping between the wave functions of $\tau$ and the neutral scalar
could be weak to make $\tau$ screened from the scalar vacuum,
resulting in an enhancement of $\tau$ couplings to charged Higgs.
For discussions in the 2HDM, see Refs.\ \cite{Andreas,Fazio,Cline,Koerner,Chen,Iguro}.
There are many other NP scenarios to explain the $R(\Ds)$ anomaly, including 
leptoquark models \cite{Dorsner,Alonso,Bauer,Barbieri,DiLuzio,Calibbi,Becirevic}, 
composite models \cite{Barbieri2,Buttazzo,Bordone,Matsuzaki},
warped extra dimensions \cite{Megias1,Megias2,DAmbrosio,Blanke0}, etc. \cite{Kang,Huang,Bardhan}.
\par
On top of the ratio $R(\Ds)$ the Belle collaboration measured the relevant polarizations 
in $B\to\Ds\tau\nu$ decays.
One can consider observable parameters associated with $D^*$ as well as $\tau$.
The $\tau$-polarization asymmetry is defined as
\begin{equation}
P_\tau(\Ds) \equiv \frac{\Gamma^{D^{(*)}}_\tau(+)-\Gamma^{D^{(*)}}_\tau(-)}
                                          {\Gamma^{D^{(*)}}_\tau(+)+\Gamma^{D^{(*)}}_\tau(-)}~,
\end{equation}
where $\Gamma^{\Ds}_\tau(\pm)$ is the decay width for $(\pm)~\tau$ helicity. 
The SM predictions are \cite{Tanaka2010,Tanaka2012}
\begin{equation}
P_\tau(D)_{\rm SM} = 0.325\pm0.009~,~~~
P_\tau(D^*)_{\rm SM} = -0.497\pm0.013~.
\end{equation}
The experimental result is \cite{Belle1612,Belle1709}
\begin{equation}
P_\tau(D^*) = -0.38\pm0.51^{+0.21}_{-0.16}~.
\end{equation}
The longitudinal $D^*$ polarization is
\begin{equation}
F_L(D^*)\equiv\frac{\Gamma(B\to D^*_L\tau\nu)}{\Gamma(B\to D^*\tau\nu)}~,
\end{equation}
where the Belle's measurement is \cite{Abdesselam}
\begin{equation}
F_L(D^*)  = 0.60\pm0.08\pm0.035~,
\end{equation}
while the SM value is estimated to be \cite{Alok2016}
\begin{equation}
F_L(D^*)_{\rm SM} = 0.46\pm0.04~.
\end{equation}
The polarization parameters could provide more information about the Lorentz structure of possible NP.
\par
In this paper we consider the nonminimal universal extra dimension (nmUED) model 
\cite{Cheng,Aguila0301,Aguila0302,Flacke08,Datta1205,Datta1408,Datta16,Datta17}
to fit the global data on $R(\Ds)$ and polarization parameters.
In the universal extra dimension (UED) models there is an extra spacelike dimension with a flat metric
compactified on an $S^1/Z_2$ orbifold, where the SM particles could reside.
Each SM particles is accompanied by infinite towers of Kaluza-Klein (KK) states.
There are two branes at the endpoints of the orbifold.
The reflection symmetry of the bulk space provides with the KK-parity conservation.
The lightest KK particle is a natural candidate for dark matter, which makes the UED scenario a strong 
alternative to the SM.
As discussed in \cite{Biswas}, in the minimal version of the UED (MUED) there are no new couplings
at the tree level relevant to $R(\Ds)$.
The radiative corrections include bulk corrections and boundary localized ones.
In the MUED models the latter is adjusted to cancel the cutoff dependent corrections.
The nmUED models allow the boundary localized terms (BLTs) to be free parameters. 
In this analysis we include the BLTs with free strength parameters.
The presence of BLTs changes mass spectrum and couplings of KK modes of the UED model. 
The NP effects enter through the possible interactions between a pair of zero-mode fermion and 
even KK-modes of charged gauge boson or scalar, associated with the BLTs \cite{Biswas,Dasgupta}.
These kinds of interactions are not allowed in the MUED because of the KK-wave function orthogonality.
Since the new interactions contribute to $R(\Ds)$ at the tree level, we expect the nmUED model would 
provide some hints to solve the $R(\Ds)$ puzzle. 

\par
The paper is organized as follows. 
In the next section the nmUED model is introduced.
Section III provides the various observables in numerical forms.
The results and discussions are given in Sec.\ IV, and we conclude in Sec.\ V.
%
\section{nmUED model}
We assume that there is one flat extra dimension ($y$) compactified on an $S^1/Z_2$ orbifold 
with radius $R$.
Two branes are located at the endpoints $y=0$ and $y=\pi R$ where both boundary terms are equal.
The 5D action for fermions $f$ is \cite{Biswas} 
\begin{eqnarray}
\calS_f&=&\sum_{f=q,\ell}\int d^4x\int_0^{\pi R}dy\left\{
    i{\bar\Psi}^f_L\Gamma^M\calD_M \Psi^f_L
+ r_f\left[\delta(y)+\delta(y-\pi R)\right]i{\bar\Psi}^f_L\gamma^\mu\calD_\mu P_L \Psi^f_L\right. \nn\\
&&\left. + i{\bar\Psi}^f_R\Gamma^M\calD_M \Psi^f_R
+ r_f\left[\delta(y)+\delta(y-\pi R)\right]i{\bar\Psi}_R\gamma^\mu\calD_\mu P_R \Psi^f_R\right\} ~,
\label{Sf}
\end{eqnarray}
where $\Psi^f_{L,R}(x,y)$ are the 5D four component Dirac spinors for fermions $f=q,\ell$.
In terms of two component spinors,
\begin{equation}
\Psi^f_{L,R}(x,y) = 
\begin{pmatrix}
\psi^f_{L,R}(x,y)\\ \chi^f_{L,R}(x,y)
\end{pmatrix}
= \sum_n
\begin{pmatrix}
\psi^{f(n)}_{L,R}(x)F^{f(n)}_{L,R}(y)\\
\chi^{f(n)}_{L,R}(x)G^{f(n)}_{L,R}(y)
\end{pmatrix}~,
\label{2compo}
\end{equation}
where $F^{f(n)}_{L,R}(y)$ and $G^{f(n)}_{L,R}(y)$ are the $n$-th KK-wave functions.
In Eq.\ (\ref{Sf}) $r_f$ is the strength of the boundary localized terms.
They are related to the mass of the $n$th KK-excitation $m_{f^{(n)}}$ by the transcendental equation
\begin{equation}
\frac{r_f m_{f^{(n)}}}{2}=\left\{
\begin{array}{cc}
-\tan\left(\frac{m_{f^{(n)}}\pi R}{2}\right)~\text{for even} ~n\\
\cot\left(\frac{m_{f^{(n)}}\pi R}{2}\right)~\text{for odd}~n
\end{array}
\right.~.
\label{TE}
\end{equation}
As for the gauge boson sector, the 5D action is
\begin{eqnarray}
\calS_{\rm gauge}&=&
-\frac{1}{4}\int d^4x\int_0^{\pi R} dy\left\{
W^i_{MN}W^{iMN} + r_V\left[\delta(y)+\delta(y-\pi R)\right]W^i_{\mu\nu}W^{i\mu\nu}\right.\nn\\
&&\left.
+B_{MN}B^{MN}+r_V\left[\delta(y)+\delta(y-\pi R)\right]B_{\mu\nu}B^{\mu\nu}\right\}~,
\end{eqnarray}
where $W^i_{MN}$, $B_{MN}$ are the 5D gauge field strength tensors.
The $n$th KK-mass of the gauge boson is
\begin{equation}
M_{W^{(n)}} = \sqrt{M_W^2+m_{V^{(n)}}^2}~,
\end{equation}
where $m_{V^{(n)}}$ satisfies the same transcendental equation as Eq.\ (\ref{TE}).
For the $5D$ scalar field $\Phi(x,y)$, the action is
\begin{equation}
\calS_\phi 
= \int d^4x\int_0^{\pi R} dy\left\{
   \left(\calD_M\Phi\right)^\dagger\left(\calD^M\Phi\right)
  +r_\phi\left[\delta(y)+\delta(y-\pi R)\right]
   \left(\calD_\mu\Phi\right)^\dagger\left(\calD^\mu\Phi\right)\right\}~.
\end{equation}
We choose $r_\phi=r_V$ for proper gauge fixing \cite{Jha}, and consequently
the mass of the KK-scalar is $m_{\phi^{(n)}}=m_{V^{(n)}}$.
The Yukawa interaction is described by
\begin{equation}
\calS_Y = -\sum_f\int d^4 x\int_0^{\pi R}dy\left\{
\lambda_5{\bar\Psi}^f_L{\tilde\Phi}\Psi^f_R
+r_Y\left[\delta(y)+\delta(y-\pi R)\right]
   \lambda_5{\bar\psi^f}_L{\tilde\Phi}\chi^f_R + {\rm H.c.}
\right\}~,
\end{equation}
where $\lambda_5$ is the $5D$ Yukawa coupling and $r_Y$ is the boundary strength.
\par
In nmUED, new KK particles contribute to $B$ decays.
As mentioned in Sec.\ I even KK-modes of $W$-boson as well as charged Higgs 
couple to a pair of zero-mode fermions, which provide new vector and scalar interactions respectively.
The effects are encoded in the overlap integrals
\begin{eqnarray}
I_n^{fg} &=& \sqrt{\pi R\left(1+\frac{r_V}{\pi R}\right)}\int_0^{\pi R} dy \Big\{
   1+r_f\left[\delta(y)+\delta(y-\pi R)\right]\Big\}a^n F_L^{f(0)}F_L^{f(0)}~,
\\\nn
I_n^{fY} &=&  \sqrt{\pi R\left(1+\frac{r_V}{\pi R}\right)}\int_0^{\pi R} dy \Big\{
   1+r_Y\left[\delta(y)+\delta(y-\pi R)\right]\Big\}f^n F_L^{f(0)}G_R^{f(0)}~,
\label{In}
\end{eqnarray}
where $a^n$ and $h^n$ are $n$th KK-mode of the $W$-boson and scalar, respectively.
For $r_\phi =r_V$, $a^n=h^n$, and further if $r_f=r_Y$ then \cite{Biswas}
\begin{equation}
I_n^{fg}=I_n^{fY}\equiv I_n^f =
\frac{\sqrt{2}(\rhat_f-\rhat_V)\sqrt{1+\rhat_V}}
		{(1+\rhat_f)\sqrt{1+r_V^2m_{V^{(n)}}^2/4+\rhat_V}}~,
\end{equation}
where $\rhat\equiv r/(\pi R)$.
Actually, $I_n^f$ is the interaction term between a pair of zero-mode fermion $f$ and 
$n$th KK-modes of $W$-boson or scalar, which encodes the NP effects on observables.
\section{Observables}
Now the effective Hamiltonian for $b\to c\ell\nu$ is 
\begin{equation}
\calH_{\rm eff}
=\frac{4G_F}{\sqrt{2}}V_{cb}\sum_{\ell=\mu,\tau}\left\{
(1+C_V^\ell)\calO_V^\ell + C_S^\ell\calO_S^\ell\right\}~,
\end{equation}
where the operators $\calO_{V,S}^\ell$ are defined by
\begin{eqnarray}
\calO_V^\ell &=& \left(\cbar_L\gamma^\mu b_L\right)\left({\bar\ell}_L\gamma_\mu\nu_{\ell L}\right)~,\\
\calO_S^\ell &=& \left(\cbar_L b_R\right) \left({\bar\ell}_R\nu_{\ell L}\right)~.
\end{eqnarray}
The NP effects are encapsulated in the Wilson coefficients $C_{V,S}^\ell$ given as \cite{Biswas}
\begin{eqnarray}
C_V^\ell &=& \sum_{n\ge 2}\left[\frac{M_W^2}{M_{W^{(n)}}^2}\right]I_n^q I_n^\ell~,\\
C_S^\ell &=& \sum_{n\ge 2}
   \left[\frac{m_b m_\ell}{M_{W^{(n)}}^2}\right]\left[\frac{M_W^2}{M_{W^{(n)}}^2}\right]
   \left\{\cos\left(\frac{1}{2}\tan^{-1}\left[\frac{m_c}{m_{f^{(n)}}}\right]
   		       -\frac{1}{2}\tan^{-1}\left[\frac{m_\ell}{m_{f^{(n)}}}\right] 
                  \right)\right.\nn\\
&&-
\left. \sin\left(\frac{1}{2}\tan^{-1}\left[\frac{m_c}{m_{f^{(n)}}}\right]
   		       +\frac{1}{2}\tan^{-1}\left[\frac{m_\ell}{m_{f^{(n)}}}\right] 
                  \right)\right\} I_n^q I_n^\ell~.
\end{eqnarray}
From $\calH_{\rm eff}$ one can calculate the transition amplitudes and decay rates for $B\to\Ds$ decays,
and construct various observable parameters.
We only concentrate on the numerical results for the observables in our analysis. 
Numerically the observables for $B\to\Ds\ell\nu_\ell$ decays are (at $\mu=m_b$ scale) \cite{Blanke}
\begin{eqnarray}
\label{obs1}
R(D) &=& 2R_{\rm SM}(D)
   \frac{\left(1+C_V^\tau\right)^2+1.54\left(1+C_V^\tau\right)C_S^\tau+1.09\left(C_S^\tau\right)^2}
      {1+\left(1+C_V^\mu\right)^2+1.54\left(1+C_V^\mu\right)C_S^\mu+1.09\left(C_S^\mu\right)^2}~,\\
R(D^*) &=& 2R_{\rm SM}(D^*)
   \frac{\left(1+C_V^\tau\right)^2+0.13\left(1+C_V^\tau\right)C_S^\tau+0.05\left(C_S^\tau\right)^2}
      {1+\left(1+C_V^\mu\right)^2+0.13\left(1+C_V^\mu\right)C_S^\mu+0.05\left(C_S^\mu\right)^2}~,\\
P_\tau(D) &=& 
 \frac{0.32\left(1+C_V^\tau\right)^2+1.54\left(1+C_V^\tau\right)C_S^\tau+1.09\left(C_S^\tau\right)^2}
 {\left(1+C_V^\tau\right)^2+1.54\left(1+C_V^\tau\right)C_S^\tau+1.09\left(C_S^\tau\right)^2}~,\\
P_\tau(D^*) &=&
\frac{-0.49\left(1+C_V^\tau\right)^2+0.13\left(1+C_V^\tau\right)C_S^\tau+0.05\left(C_S^\tau\right)^2}
{\left(1+C_V^\tau\right)^2+0.13\left(1+C_V^\tau\right)C_S^\tau+0.05\left(C_S^\tau\right)^2}~,\\
F_L(D^*) &=&
\frac{0.46\left(1+C_V^\tau\right)^2+0.13\left(1+C_V^\tau\right)C_S^\tau+0.05\left(C_S^\tau\right)^2}
{\left(1+C_V^\tau\right)^2+0.13\left(1+C_V^\tau\right)C_S^\tau+0.05\left(C_S^\tau\right)^2}~,\\
{\rm Br}(B_c\to\tau\nu) &=&
   0.02\left(\frac{f_{B_c}}{0.43~{\rm GeV}}\right)\Big[1+C_V^\tau+4.3C_S^\tau\Big]^2~.
\label{obs5}
\end{eqnarray}
The results are obtained from the numerical values of the relevant form factors of $B\to D$ \cite{Aoki} 
and $B\to D^*$ transitions \cite{Amhis,Bernlochner}.
\par
The branching ratio of $B_c\to\tau\nu$, ${\rm Br}(B_c\to\tau\nu)$ could impose strong constraints 
on $R(\Ds)$ \cite{Alonso2016}.
Since ${\rm Br}(B_c\to\tau\nu)\sim\left(1+C_V^\tau+4.3C_S^\tau\right)^2$, 
the branching ratio directly affects the relevant Wilson coefficients.
There are still debates on the upper bound of ${\rm Br}(B_c\to\tau\nu)$.
The strongest bound is from Ref.\ \cite{Akeroyd} where ${\rm Br}(B_c\to\tau\nu)<10\%$.
On the other hand, Ref.\ \cite{Blanke} argues that the branching ratio could be as large as $60\%$.
In this analysis we do not explicitly impose the ${\rm Br}(B_c\to\tau\nu)$ constraints,
because as we will see later our results are compatible with small values of ${\rm Br}(B_c\to\tau\nu)$.
The experimental data for various observables used in this analysis are listed in Table \ref{T1}.
%
\begin{table}
\begin{tabular}{|c|| cc |}\hline
              & ~$R(D)$ & ~$R(D^*)$   \\\hline
 BABAR & ~$0.440\pm0.058\pm0.042$ & ~$0.332\pm0.024\pm0.018$ \cite{BaBar1} \\
 Belle(2015) & ~$0.375\pm0.064\pm0.026$ & ~$0.293\pm0.038\pm0.015$ \cite{Belle1} \\
 Belle(2016) & ~$-$ & ~$0.302\pm0.030\pm0.011$ \cite{Belle1607} \\
 Belle(2017) & ~$-$ & ~$0.276\pm0.034^{+0.029}_{-0.026}$ \cite{Belle1703} \\
 Belle(2017) & ~$-$ & ~$0.270\pm0.035^{+0.028}_{-0.025}$ \cite{Belle1612,Belle1709} \\
 Belle(2019) & ~$0.307\pm0.037\pm0.016$ & ~$0.283\pm0.018\pm0.014$ \cite{Belle1904}\\
 LHCb(2015) & ~$-$ & ~$0.336\pm0.027\pm0.030$ \cite{LHCb1} \\
 LHCb(2017) & ~$-$ & ~$0.291\pm0.019\pm0.026\pm0.013$ \cite{LHCb2} \\\hline\hline
 & $P_\tau(D^*)$ & $F_L(D^*)$ \\\hline
 Belle(2017) & $-0.38\pm0.51^{+0.21}_{-0.16}$\cite{Belle1612,Belle1709} & $-$ \\
 Belle(2019)  & $-$ & $0.60\pm0.08\pm0.04$ \cite{Abdesselam} \\
 \hline
 \end{tabular}
\caption{Experimental data for $R(\Ds)$, $P_\tau(\Ds)$ and $F_L(D^*)$.
The uncertainties are $\pm$(statistical)$\pm$(systematic).
For the third uncertainty of LHCb(1711), see \cite{LHCb2} for details.
For BABAR, Belle(2015), and Belle(2019) results, the correlations between $R(D)$ and $R(D^*)$ are
$-0.31$, $-0.50$ and $-0.51$ respectively \cite{HFAG2019}.}
\label{T1}
\end{table}
%
\section{Results}
We implement the global $\chi^2$ fit for the observables in Table \ref{T1}.
We first define the $\chi^2$ as
\begin{equation}
\chi^2\equiv\sum_{i,j}
\left[ \calO_i^{\rm exp}-\calO_i^{\rm th}\right]\calC_{ij}^{-1}
\left[\calO_j^{\rm exp}-\calO_j^{\rm th}\right]~,
\label{chisq}
\end{equation}
where $\calO_i^{\rm exp}$ are the experimental data 
while $\calO_i^{\rm th}$ are the theoretical predictions of Eqs.(\ref{obs1})-(\ref{obs5}),
and $\calC_{ij}$ are the correlation matrix elements.
\par
There are two major constraints.
One is from the oblique parameters of the electroweak precision test  (EWPT) \cite{Flacke2012,Flacke2013,Datta2013,Dey}.
In the nmUED model, the Fermi constant is modified by the tree level contributions of even $n$th
KK-modes of $W$-bosons to the four-fermion interactions.
This kind of correction is absent in the MUED scenario.
The Fermi constant in nmUED is now written as 
\begin{equation}
G_F = G_F^0 + \delta G_F~.
\end{equation}
Here $G_F^0$ is the Fermi constant in the SM and $\delta G_F$ is the correction from the new 
contributions of $W^\pm$ KK-modes.
Explicitly \cite{Biswas},
\begin{equation}
G_F^0 = \frac{g^2}{4\sqrt{2}M_W^2}~,~~~
\delta G_F = \sum_{n\ge 2}\frac{g^2 (I_n^\ell)^2}{4\sqrt{2}m_{W^{(n)}}^2}~,
\end{equation}
where $g$ is the gauge coupling constant.
Note that $\delta G_F\sim (I_n^\ell)^2$ because the Fermi constant is derived from the muon lifetime.
We only consider the 2nd KK contributions for simplicity.
Now the Fermi constant is related to the Peskin-Tacheuchi parameters as \cite{Flacke2012}
\begin{equation}
S_{\rm nmUED} = 0~,~~~
T_{\rm nmUED} = -\frac{1}{\alpha}\frac{\delta G_F}{G_F}~,~~~
U_{\rm nmUED} = \frac{4\sin^2\theta_W}{\alpha}\frac{\delta G_F}{G_F}~,
\end{equation}
where we neglect possible loop effects which are subdominant compared to the tree-level contributions 
to $\delta G_F$.
We use the data \cite{Gfitter}
\begin{equation}
S=0.05\pm0.11~,~~~T=0.09\pm 0.13~,~~~U=0.01\pm 0.11~,
\label{TUdata}
\end{equation}
where the correlation coefficients are 
\begin{equation}
\rho_{ST} = 0.90~,~~~\rho_{TU} = -0.83~,~~~\rho_{US} = -0.59~.
\end{equation}
Following the methods of \cite{Dey}, we impose the $S$, $T$, $U$ constraints by requiring 
$\chi_{STU}^2<6.18$ at $2\sigma$ where $\chi_{STU}^2$ is defined by the covariant matrix relevant for
the $S$, $T$, $U$ parameters, similarly to Eq.\ (\ref{chisq}).
\par
The other major constraint comes from the LHC dilepton resonance searches.
At the LHC the second KK gauge boson $A^{(2)}$ can be produced via the KK number violating interactions,
subsequently decaying into the SM particles.
Recent results from ATLAS dilepton resonance searches at the 13 TeV with $13.3~{\rm fb}^{-1}$
provide a stringent constraint on the nmUED parameters \cite{Flacke2017}.
We reflect the results of \cite{Flacke2017} on the strength of the BLKT in the gauge sector
to constrain our analysis to the region $0\le r_V/R \le 0.5$.
The best-fit values for the minimum $\chi^2$ are listed in Table \ref{T2}.
\begin{table}
\begin{tabular}{ccccccc}
$R(D)$ & $R(D^*)$ & $P_\tau(D)$ & $P_\tau(D^*)$ & $F_L(D^*)$ & ${\rm Br}(B_c\to\tau\nu)$ 
& $\chi^2_{\rm min}/{\rm d.o.f.}$ \\\hline
 $0.343$ & $0.296$ & $0.320$ & $-0.490$ & $0.460$ & $2.75\times 10^{-2}$ & $1.25$ \\\hline
\end{tabular}
\caption{Best-fit values.}
\label{T2}
\end{table}
\par
\begin{figure}
\begin{tabular}{cc}
\includegraphics[scale=0.12]{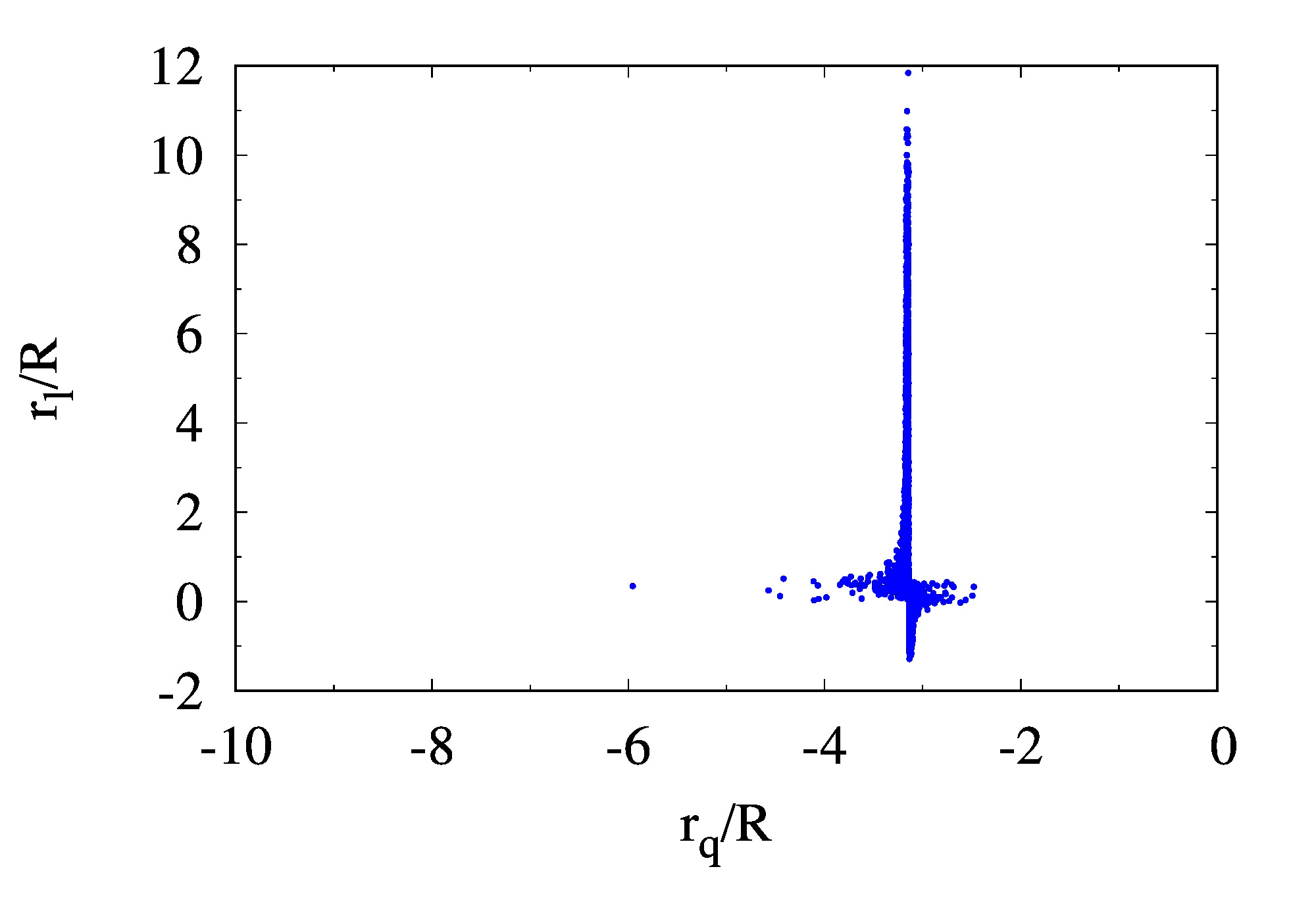}
&\includegraphics[scale=0.12]{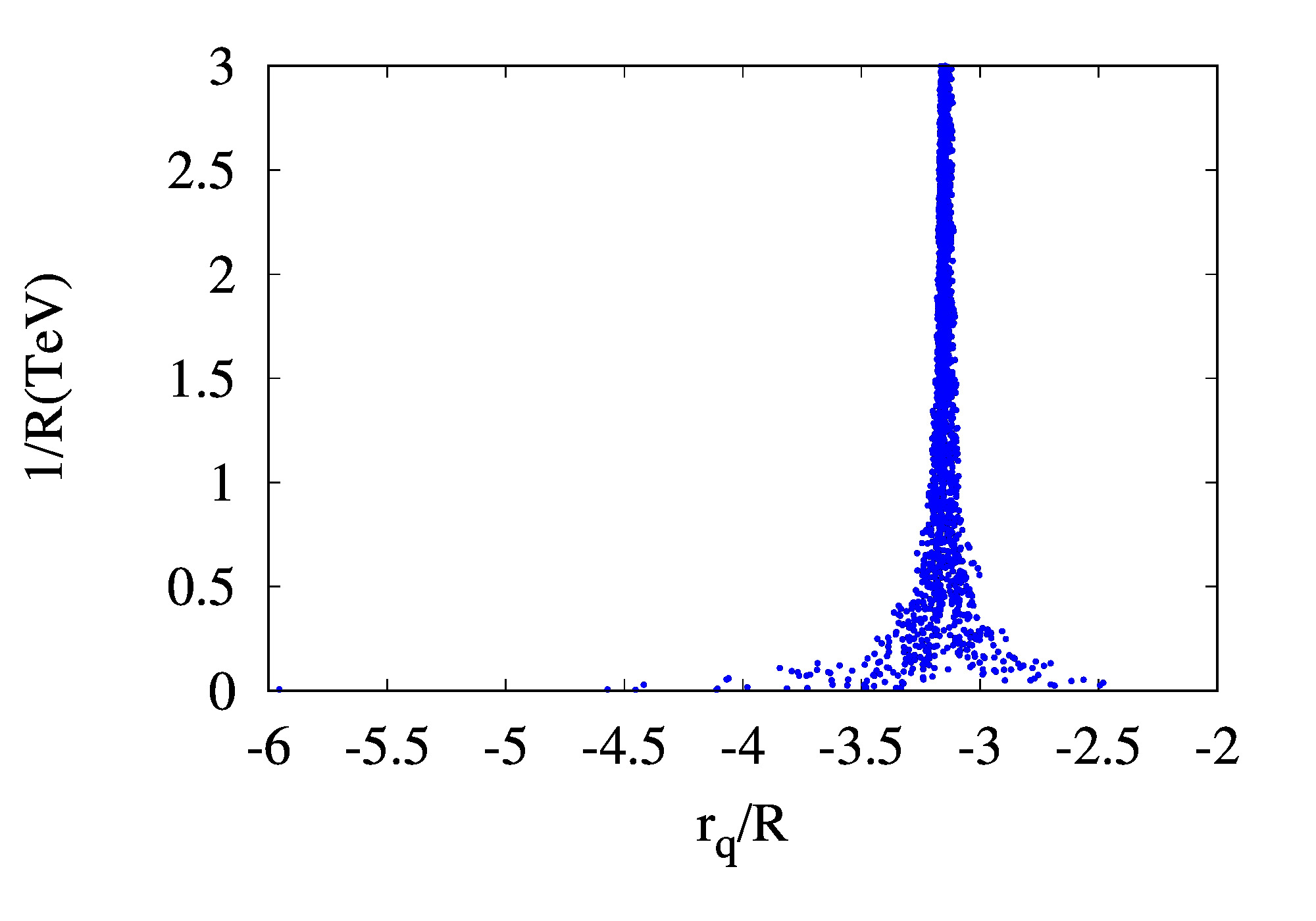}\\
(a) & (b)\\
\includegraphics[scale=0.12]{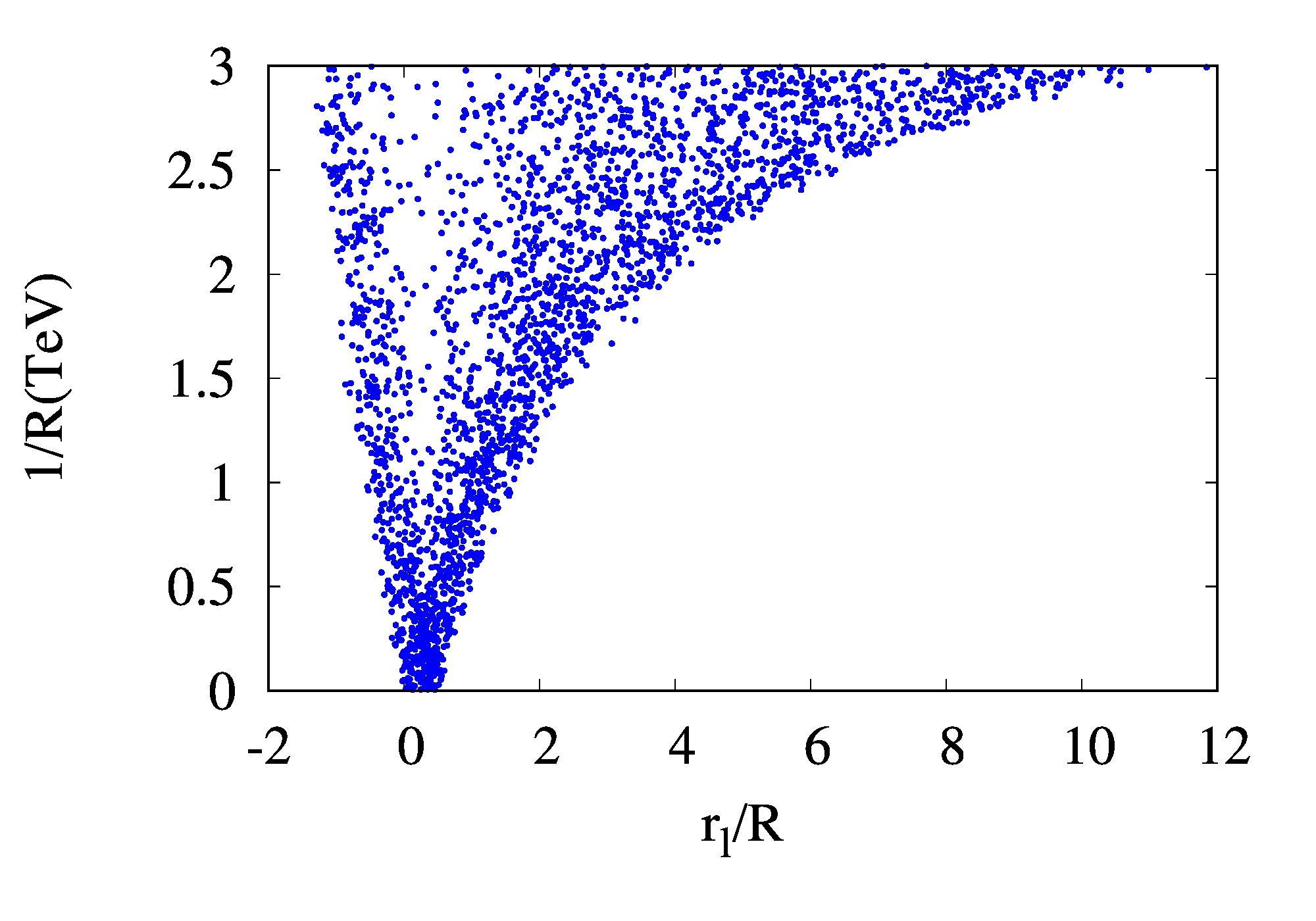}
&\includegraphics[scale=0.12]{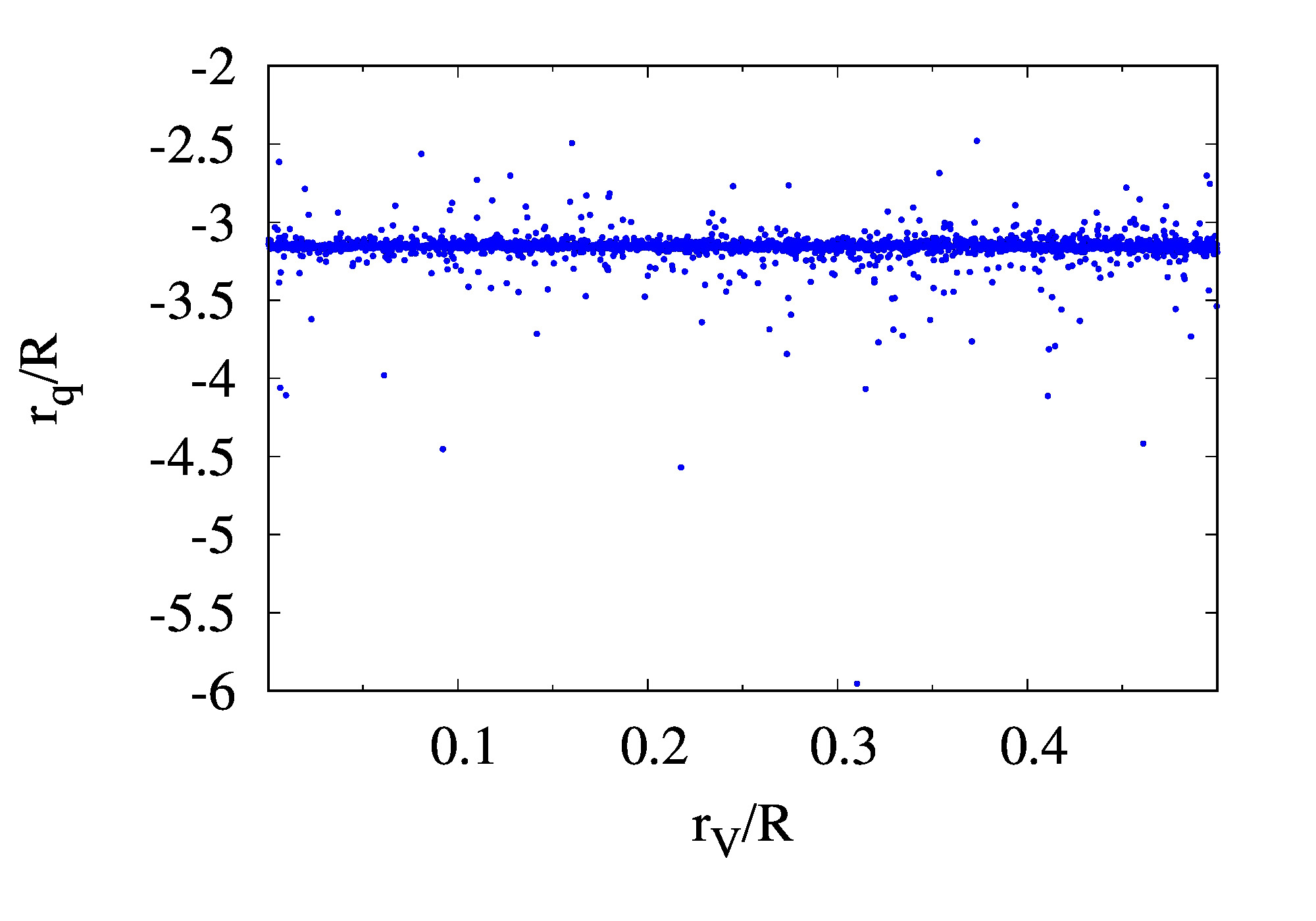}\\
(c) & (d)
\end{tabular}
\caption{\label{F1}Allowed regions of model parameters of nmUED at the $2\sigma$ level.}
\end{figure}

In Fig.\ \ref{F1}, we plot the allowed regions of the nmUED parameters at the $2\sigma$ level.
We scanned over the range $0\le 1/R \le 3~{\rm TeV}$.
A noticeable feature is that the allowed range of $r_q/R$ is rather narrow with negative values, 
contrary to that of $r_\ell/R$ as shown in Fig.\ \ref{F1} (a).
In the nmUED models, $r_{q,\ell}$ are considered as free parameters and can be negative
but with some restrictions.
The fields $F_{L,R}^{f(n)}$ and $G_{L,R}^{f(n)}$ of Eq.\ (\ref{2compo}) have normalization factor 
\cite{Datta16,Datta17,Biswas},
\begin{equation}
N_n^f =
 \sqrt{\frac{2}{\pi R}}
 \frac{1}{\sqrt{1+r_f^2m_{f^{(n)}}^2/4+r_f/(\pi R)}}~.
\label{Nnf}
\end{equation}
For $N_n^f$ to be meaningful, 
$r_f/(\pi R) > -1-r_f^2 m_{f^{(n)}}^2/4$ and for small values of $r_f m_{f^{(n)}}$, 
$r_f/R\gtrsim -\pi$. 
Our results of Fig.\ \ref{F1} satisfy this requirement.
Note that the points near $r_q/R=-\pi $ are favorable for larger $R(D)$ and smaller $\chi^2$.
\par
\begin{figure}
\begin{tabular}{cc}
\includegraphics[scale=0.13]{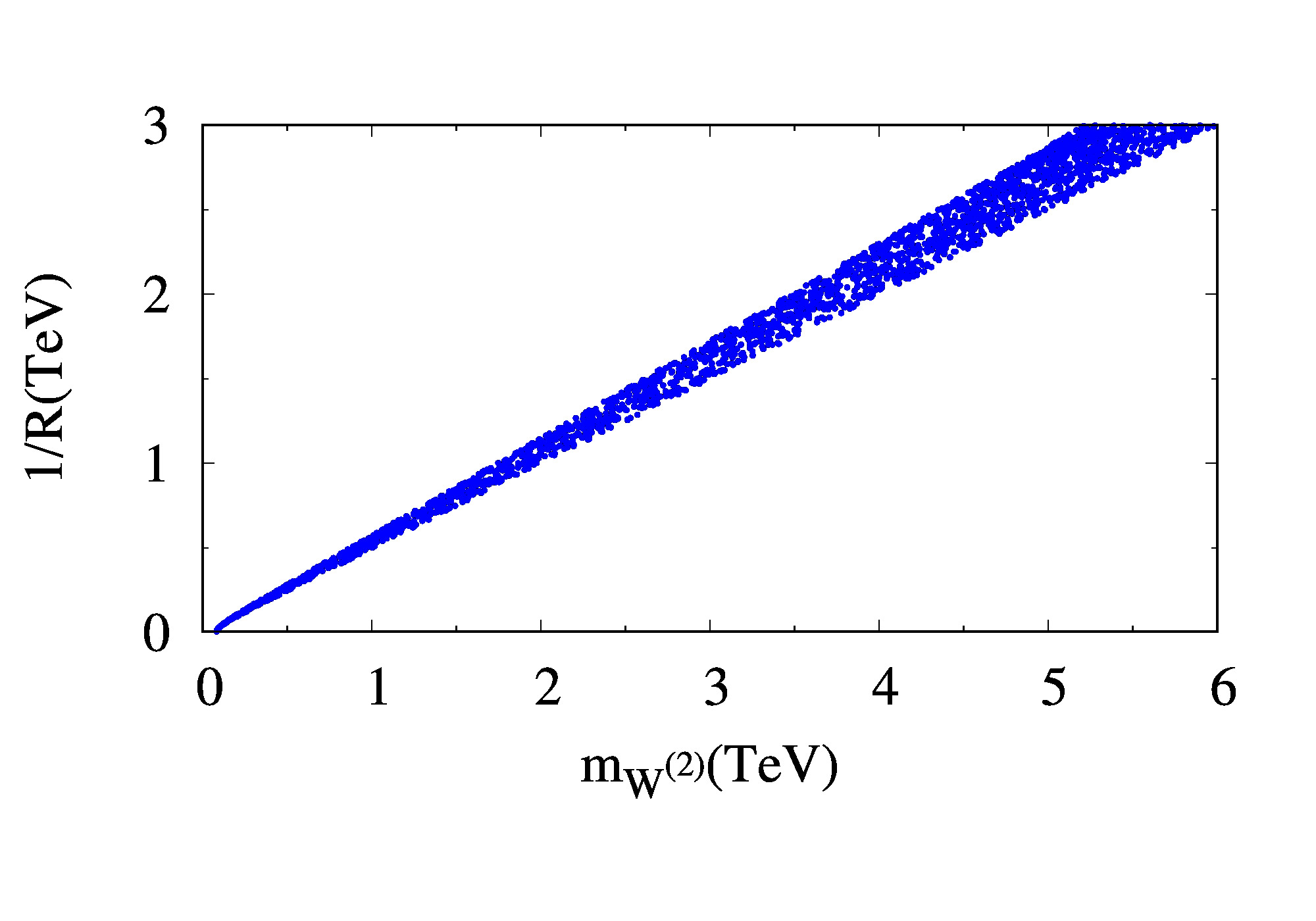}
&\includegraphics[scale=0.13]{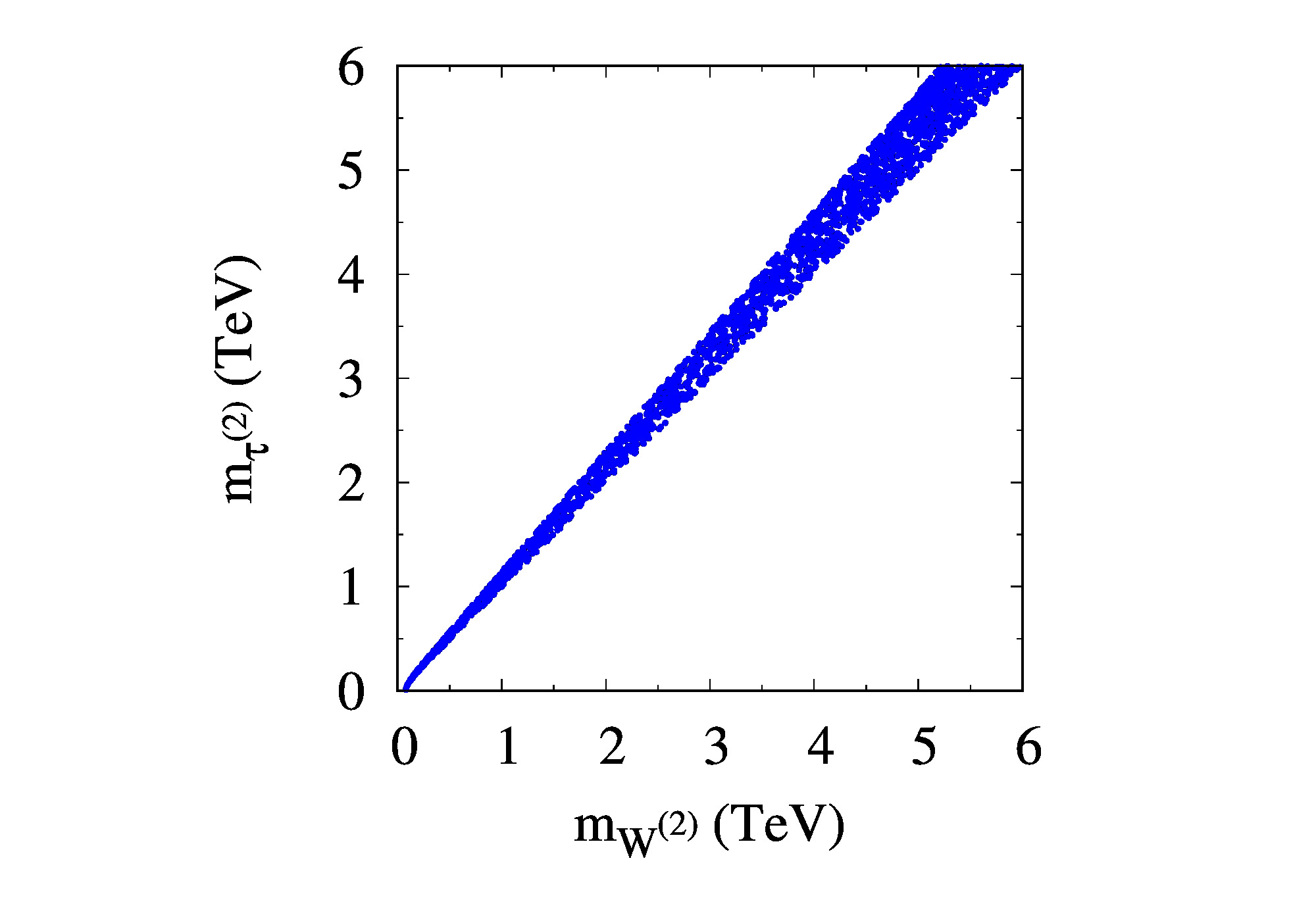}\\
(a) & (b) 
\end{tabular}
\caption{\label{F2}Mass scales of nmUED at $2\sigma$.}
\end{figure}
\begin{figure}
\begin{tabular}{cc}
\includegraphics[scale=0.12]{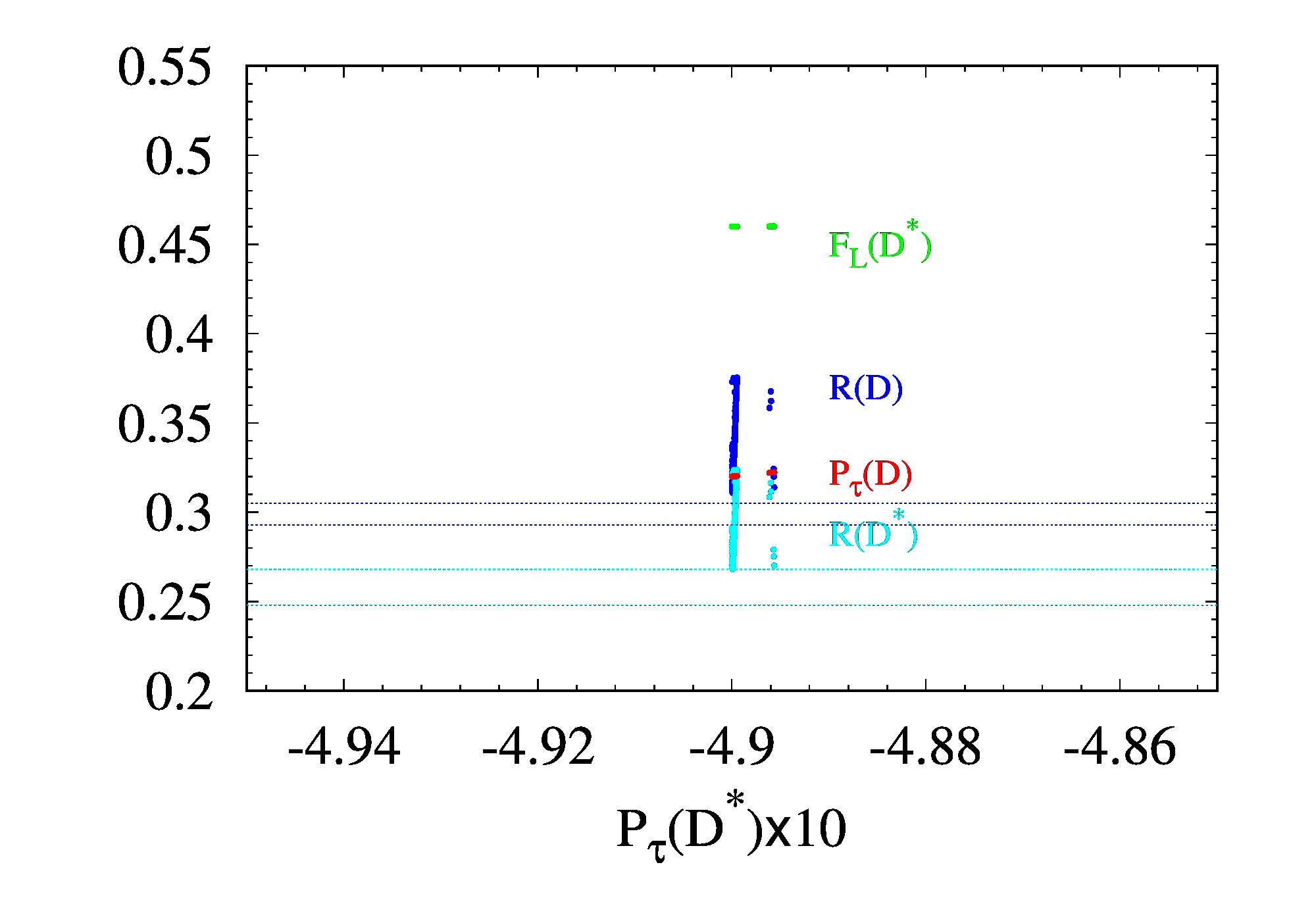}
&\includegraphics[scale=0.12]{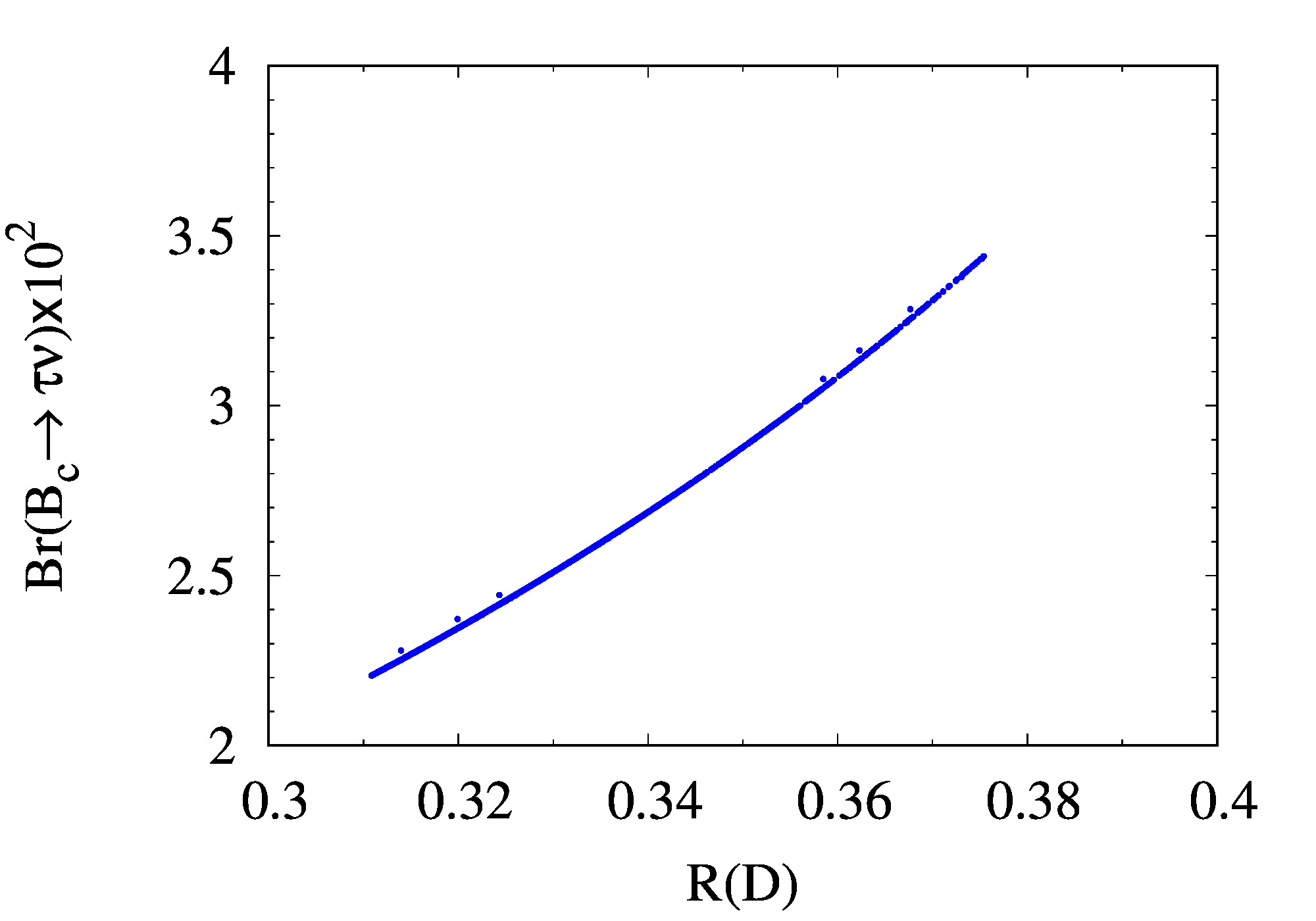}\\
(a) & (b) 
\end{tabular}
\caption{\label{F_obs}Allowed values for various observables at $2\sigma$. 
In (a) numerical values of $R(\Ds)$, $P_\tau(D)$, and $F_L(D^*)$ are plotted with respect to the 
values of $P_\tau(D^*)$.
The horizontal lines are the SM predictions at $2\sigma$ for 
$R(D)$ (blue) and $R(D^*)$ (cyan).
Other polarization parameters $P_\tau(\Ds)$ and $F_L(D^*)$ are consistent with the SM values
at $2\sigma$.
In (b) the branching ratio of ${\rm Br}(B_c\to\tau\nu)$ vs $R(D)$ is plotted.}
\end{figure}

Figure \ref{F2} shows the 2nd KK masses $m_{W^{(2)}}$ and $m_{\tau^{(2)}}$.
Allowed values of various observables at $2\sigma$ are given in Fig.\ \ref{F_obs}.
As can be seen in Fig.\ \ref{F_obs} (a), $R(D)$ is still far away from the SM predictions
beyond $2\sigma$ level while $R(D^*)$ values have small overlaps at the edge of the SM-allowed range  within $2\sigma$ .
But the best-fit values of $R(\Ds)$ in Table \ref{T1} are still beyond the SM by more than $2\sigma$.
Other polarization observables $P_\tau(\Ds)$ and $F_L(D^*)$ are consistent with the SM.
Figure \ref{F_obs} (b) shows that the branching ratio ${\rm Br}(B_c\to\tau\nu)$ lies safely 
within a few percents. 
\par
\begin{figure}
\begin{tabular}{cc}
\includegraphics[scale=0.12]{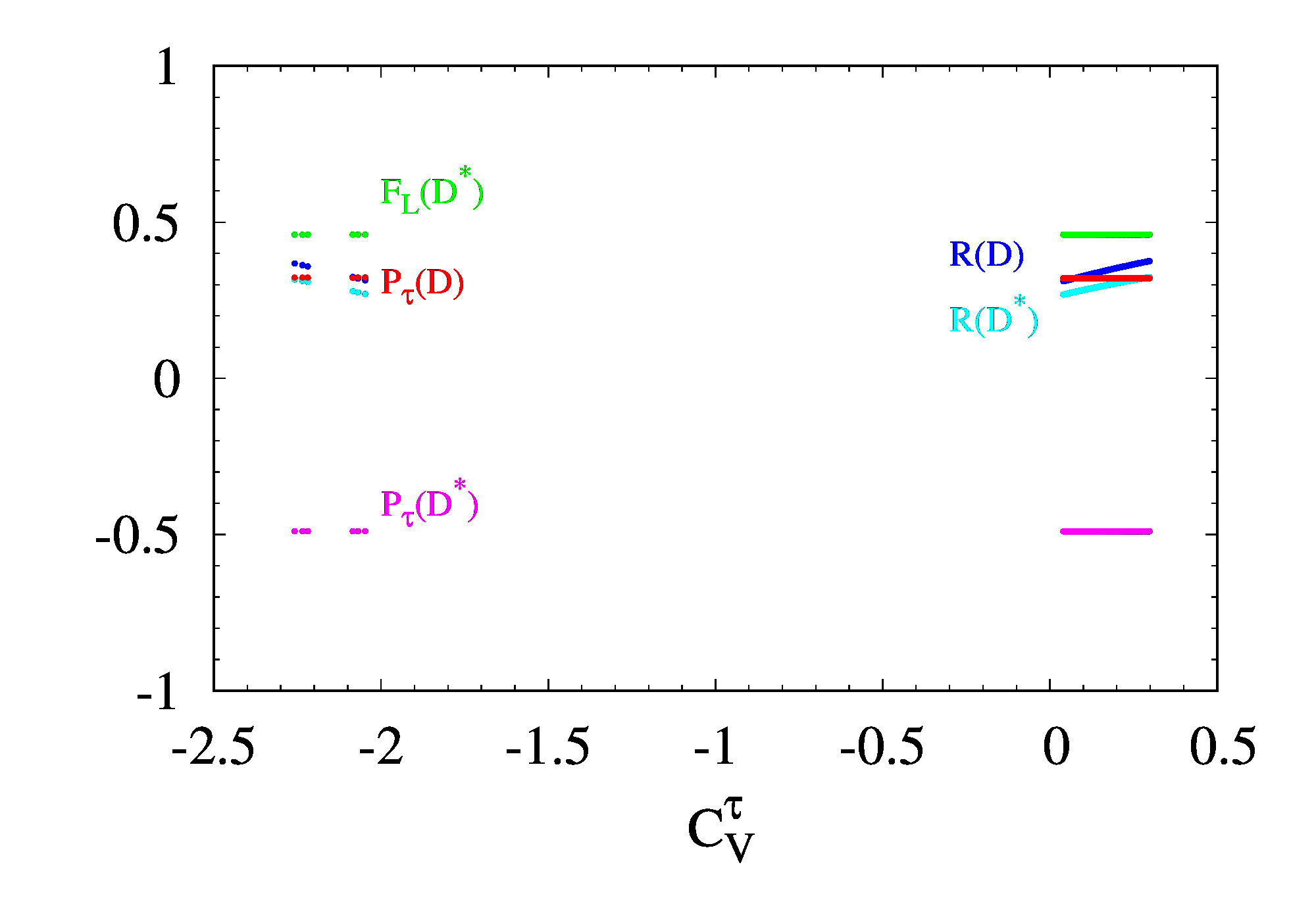}
&\includegraphics[scale=0.12]{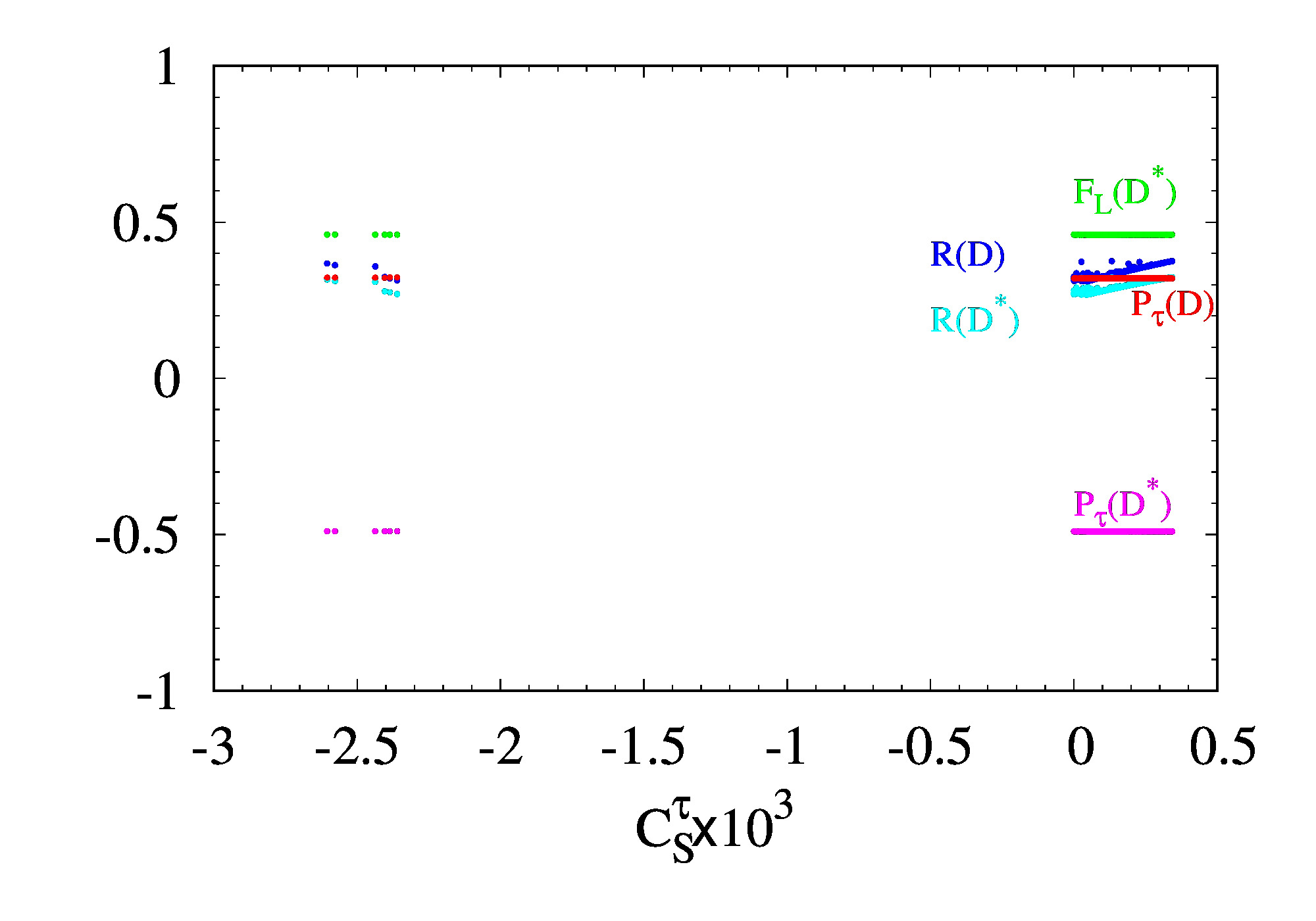}\\
(a) & (b)
\end{tabular}
\caption{\label{FWO}Wilson coefficients $C_{V,S}^\tau$ and observables.}
\end{figure}
\begin{figure}
\begin{tabular}{c}
\includegraphics[scale=0.2]{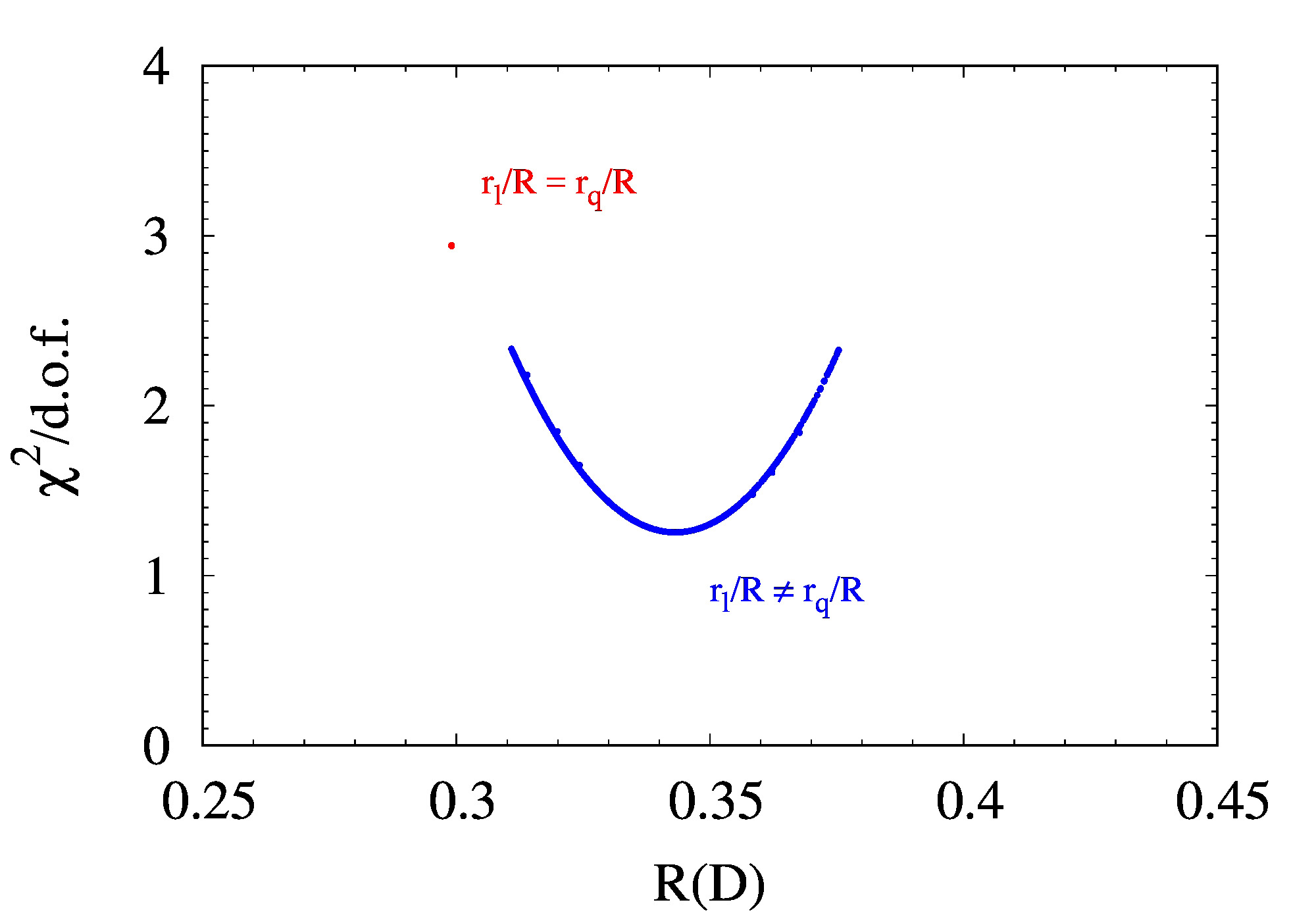}
\end{tabular}
\caption{\label{F_chisq}$R(D)$ vs $\chi^2/{\rm d.o.f.}$ for $r_\ell/R=r_q/R$ and $r_\ell/R\ne r_q/R$.}
\end{figure}

Contributions of the Wilson coefficients to observables at $2\sigma$ are depicted in Fig.\ \ref{FWO}.
We find that the pattern for $C_V^\mu$ is very similar to that of $C_S^\mu$.
Note that the Wilson coefficients are
\begin{equation}
C_{V,S}\sim I_n^q I_n^\ell~,
\end{equation}
while the EW precision parameters are
\begin{equation}
T_{\rm nmUED}(U_{\rm nmUED})\sim\delta G_F\sim \left(I_n^\ell\right)^2~.
\label{TU}
\end{equation}
In case of $r_q = r_\ell$ the overlap integrals become $I_n^q=I_n^\ell$ and $C_{V,S}\sim\left(I_n^\ell\right)^2$,
which are directly affected by the oblique parameters of Eq.\ (\ref{TU}). 
According to Eq.\ (\ref{TUdata}) EWPT prefers small $\left(I_n^\ell\right)^2$.
It means that for $r_q=r_\ell$ EWPT requires smaller $C_{V,S}$,  which results in smaller $R(D^{(*)})$ and 
does not fit data so well.
In other words, we find that $R(D^{(*)})$ anomalies require $r_q\ne r_\ell$ in nmUED.
The situation is depicted in Fig.\ \ref{F_chisq} where $R(D)$ vs $\chi^2/{\rm d.o.f.}$ are compared for
$r_\ell/R=r_q/R$ and $r_\ell/R\ne r_q/R$ cases.
\par
\par
\begin{figure}
\begin{tabular}{cc}
\includegraphics[scale=0.12]{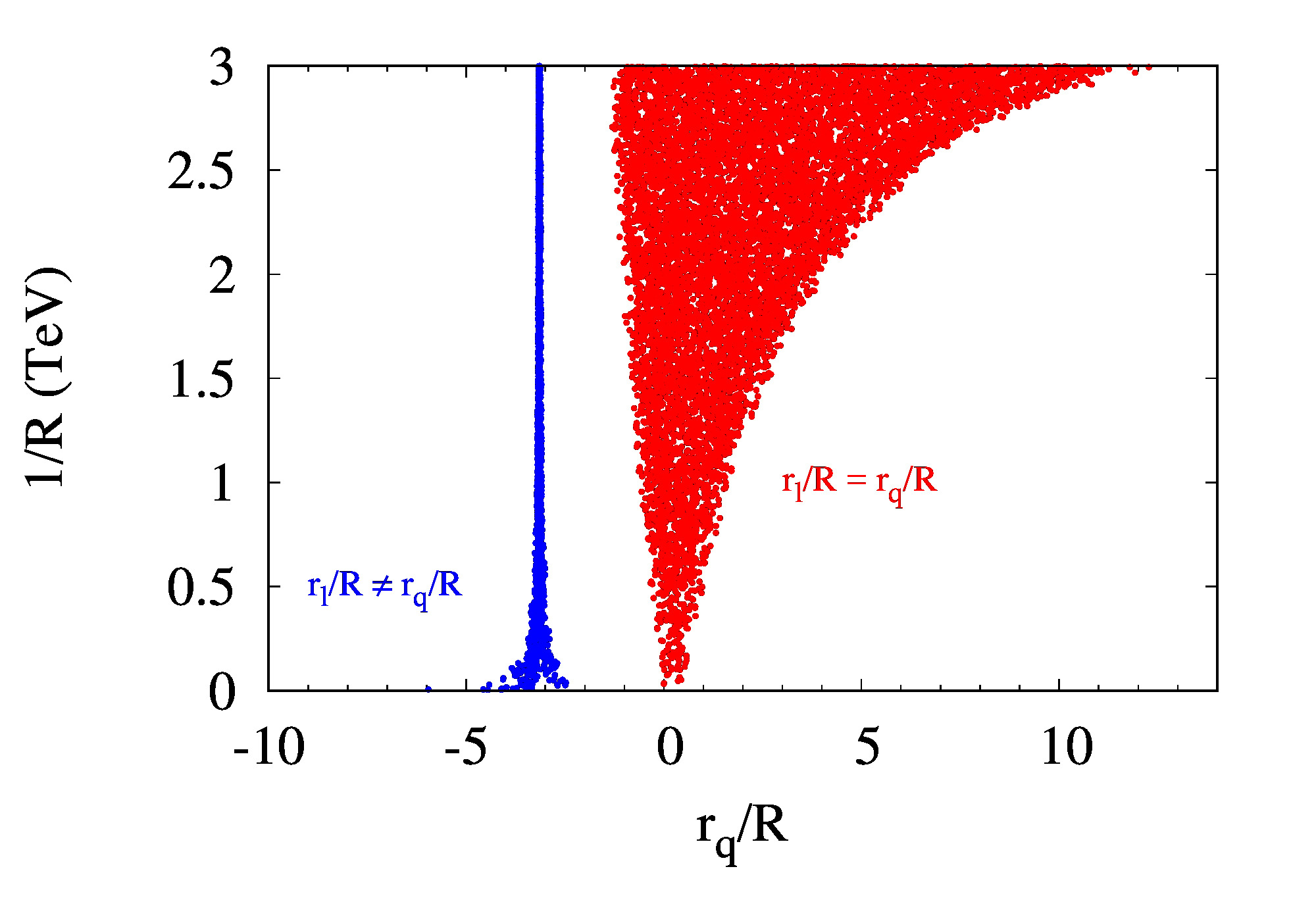} &
\includegraphics[scale=0.12]{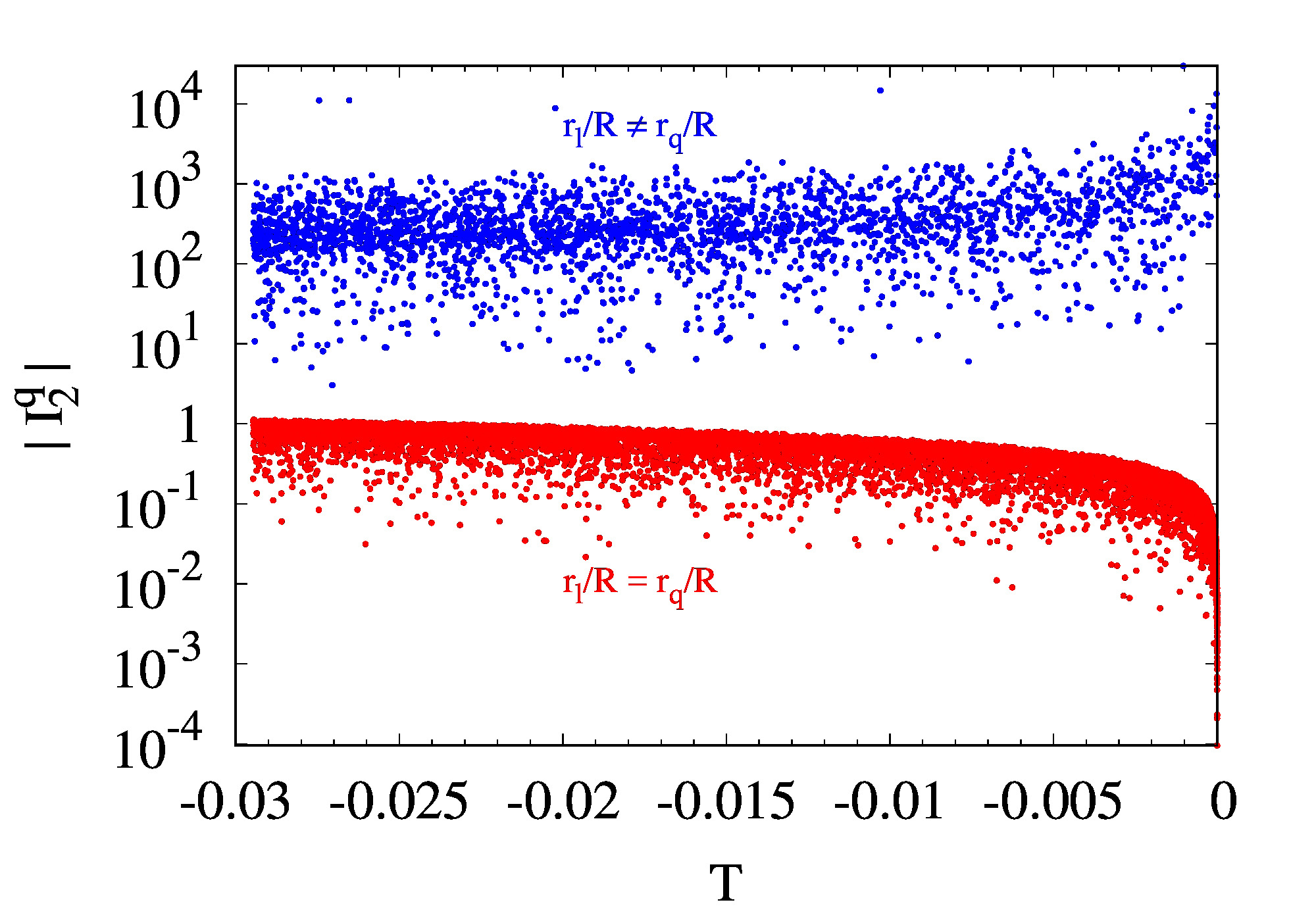} \\
(a) & (b) \\
\includegraphics[scale=0.12]{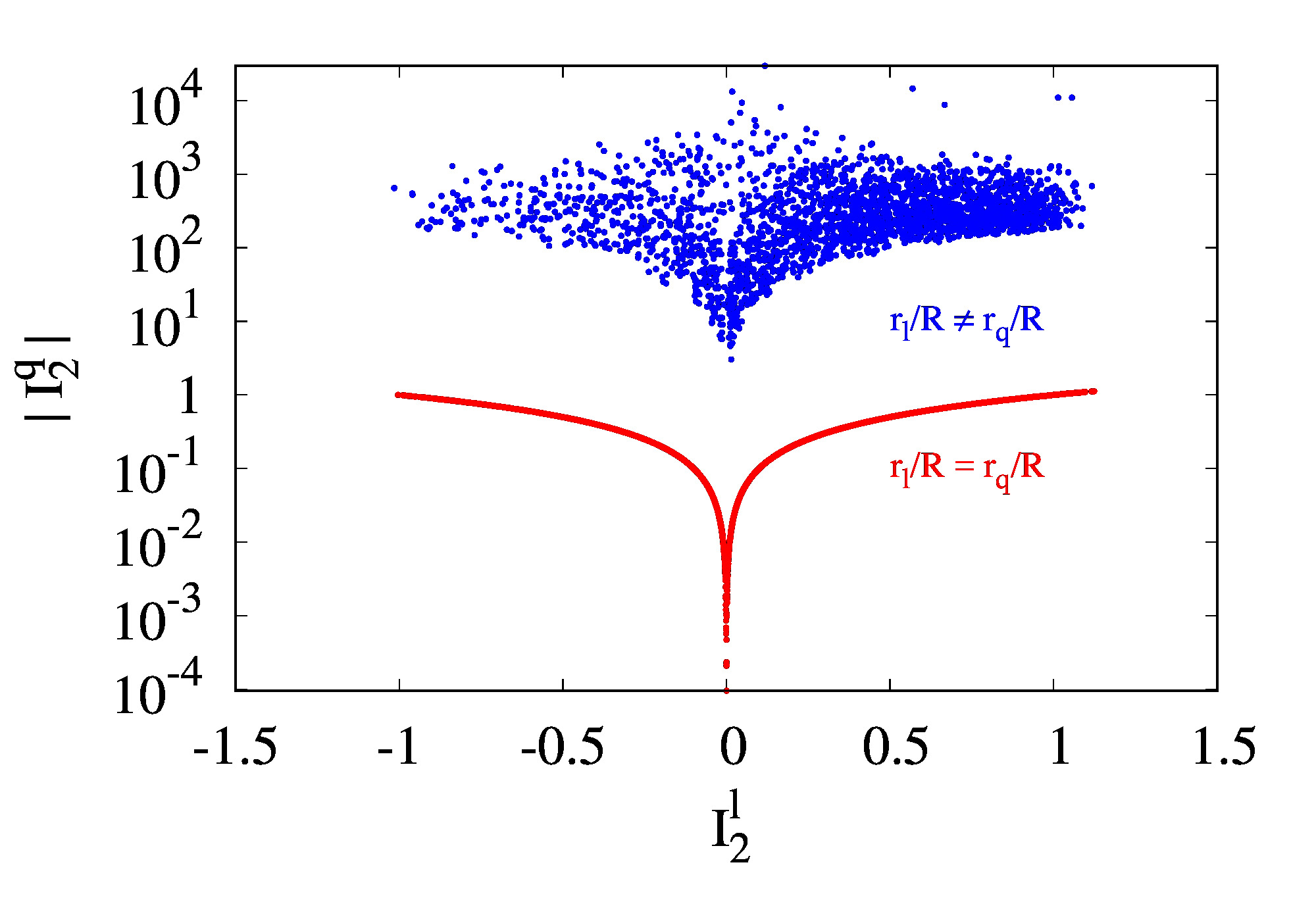} &
\includegraphics[scale=0.12]{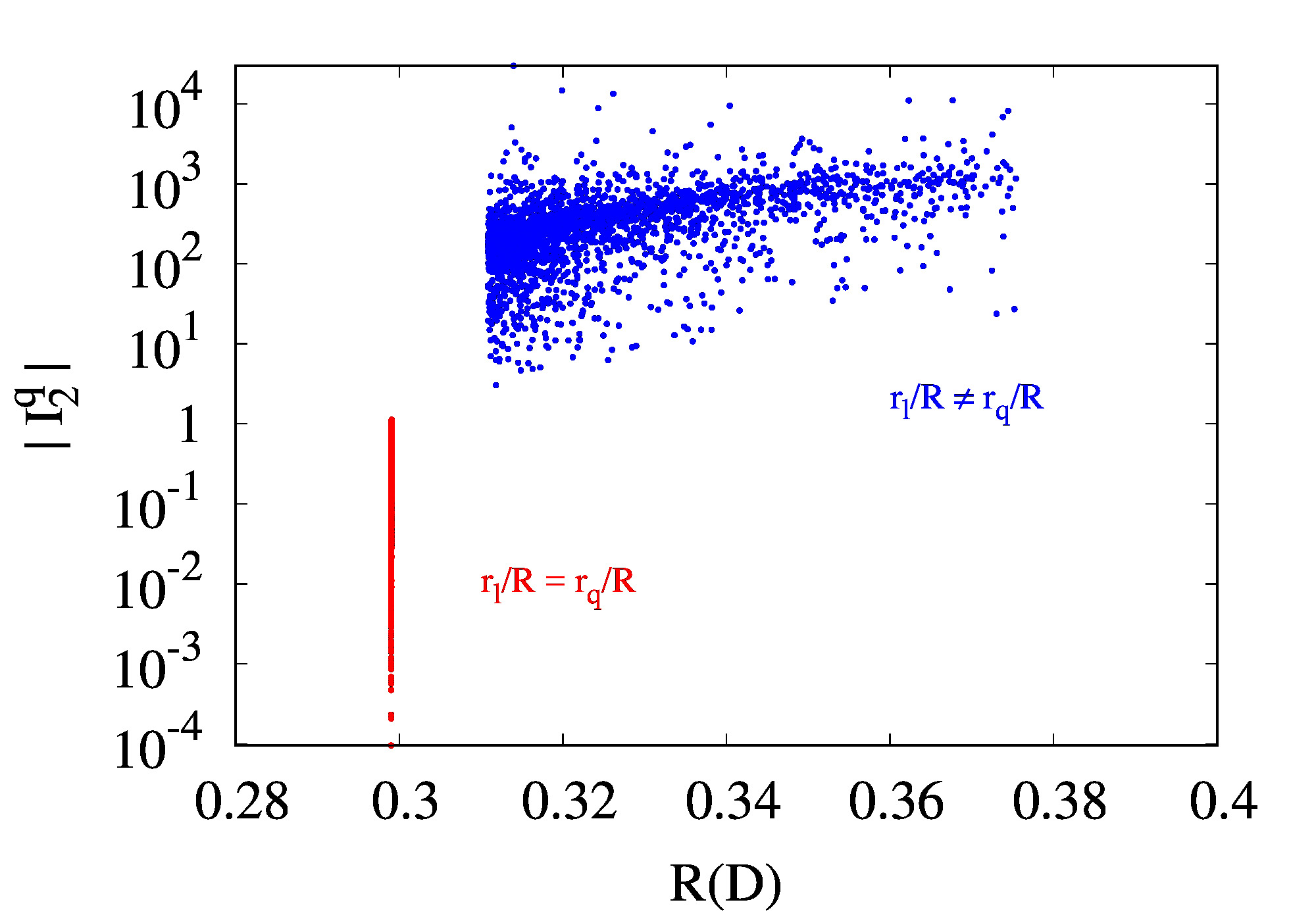} \\
(c) & (d)
\end{tabular}\caption{\label{noRrl}
Comparisons of various parameter spaces for $r_q/R = r_\ell/R$ (red) and $r_q/R\ne r_\ell/R$ (blue)
at $2\sigma$.}
\end{figure}

To see the effects of $r_\ell/R\ne r_q/R$ more dramatically,
we compare the cases of $r_\ell/R=r_q/R$ and $r_\ell/R\ne r_q/R$ in Fig.\ \ref{noRrl}.
Figure \ref{noRrl} (a) shows that the allowed regions of $r_q/R$ are quite different from each other. 
The effect of $r_\ell/R\ne r_q/R$ appears dramatically on $I_2^q$, as shown in Figs.\ \ref{noRrl} (b)-(d).
As mentioned above, this is due to the constraints on the oblique parameters.
If $r_q/R=r_\ell/R$, then $I_2^q=I_2^\ell$ and it should be kept small to satisfy the EWPT 
(Fig.\ \ref{noRrl} (b)).
In case of $r_q/R\ne r_\ell/R$, $I_2^q$ can be very large compared to $I_2^\ell$ (Fig.\ \ref{noRrl} (c)).
As a result, $R(D)$ is allowed to have large values to fit the data (Fig.\ \ref{noRrl} (d)).
\par
\begin{figure}
\begin{tabular}{cc}
\includegraphics[scale=0.12]{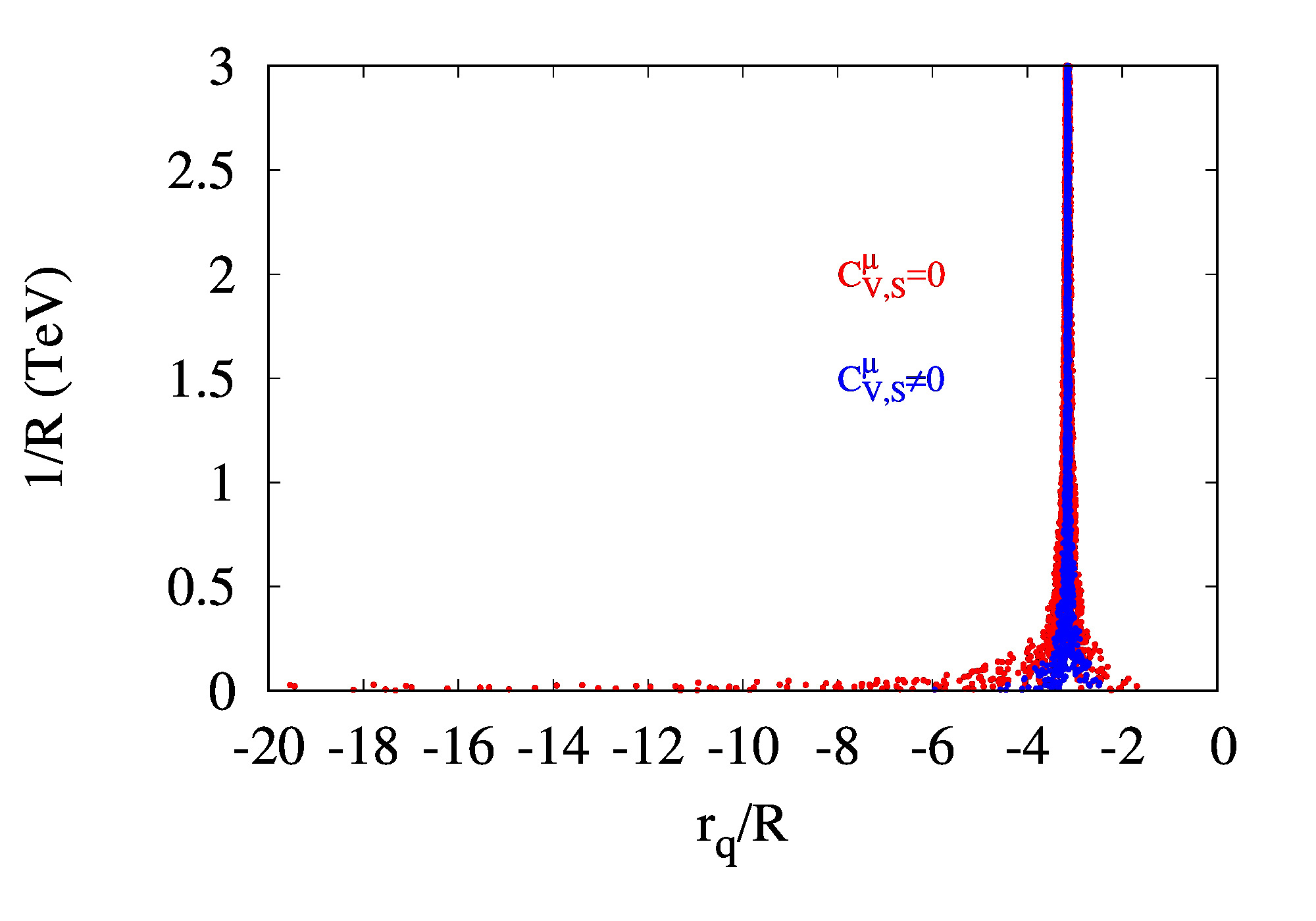} &
\includegraphics[scale=0.12]{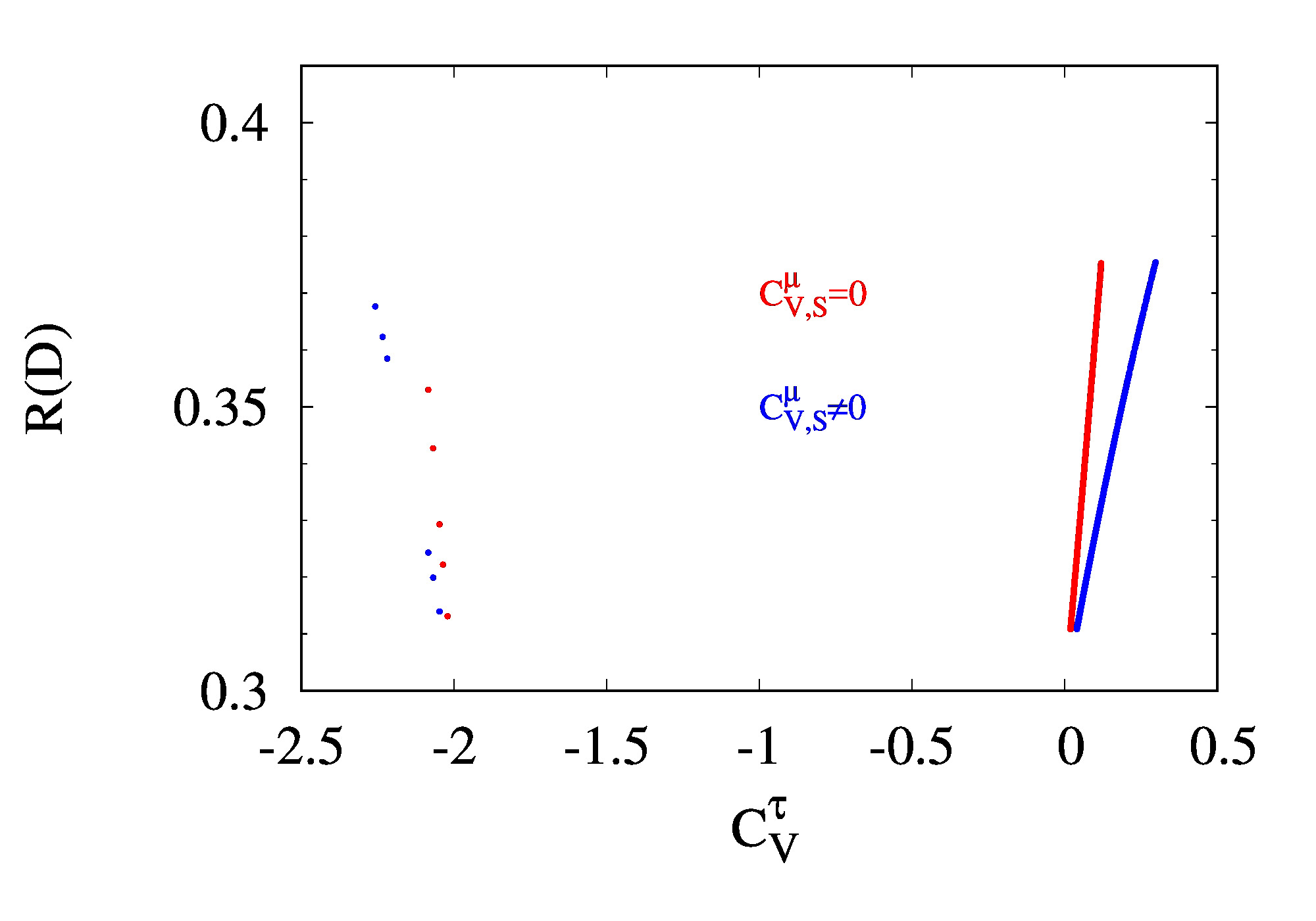} \\
(a) & (b) 
\end{tabular}\caption{\label{noCmu}
Comparisons of parameter spaces for $C_{V,S}^\mu=0$ (red) and $C_{V,S}^\mu\ne 0$ (blue)
at $2\sigma$.}
\end{figure}

In our analysis $C_V^\tau = C_V^\mu$, and we checked the influence of nonzero $C_{V,S}^\mu$.
Figure \ref{noCmu} shows some of the results.
Figure \ref{noCmu} (a) depicts $1/R$ vs $r_q/R$ while (b) does $R(D)$ vs $C_V^\tau$.
We have similar figure for $R(D^*)$ to Fig.\ \ref{noCmu} (b). 
Whether $C_{V,S}^\mu =0$ or not does not affect the observables including the polarizations so much, 
but the allowed range of $r_q/R$ or $C_V^\tau$ could be slightly different. 
The effect of $C_S^\mu$ is negligible because its values are very small compared to $C_S^\tau$.
Note that $C_S^\mu$ is suppressed by $\sim m_\mu/m_\tau$ with respect to $C_S^\tau$.
And the mixed terms of $(1+C_V^{\mu,\tau})C_S^{\mu,\tau}$ in Eq.\ (\ref{obs1}) are the main source 
of a difference between $\tau\nu$ mode and $\mu\nu$ mode.
\par
\begin{figure}
\begin{tabular}{cc}
\includegraphics[scale=0.12]{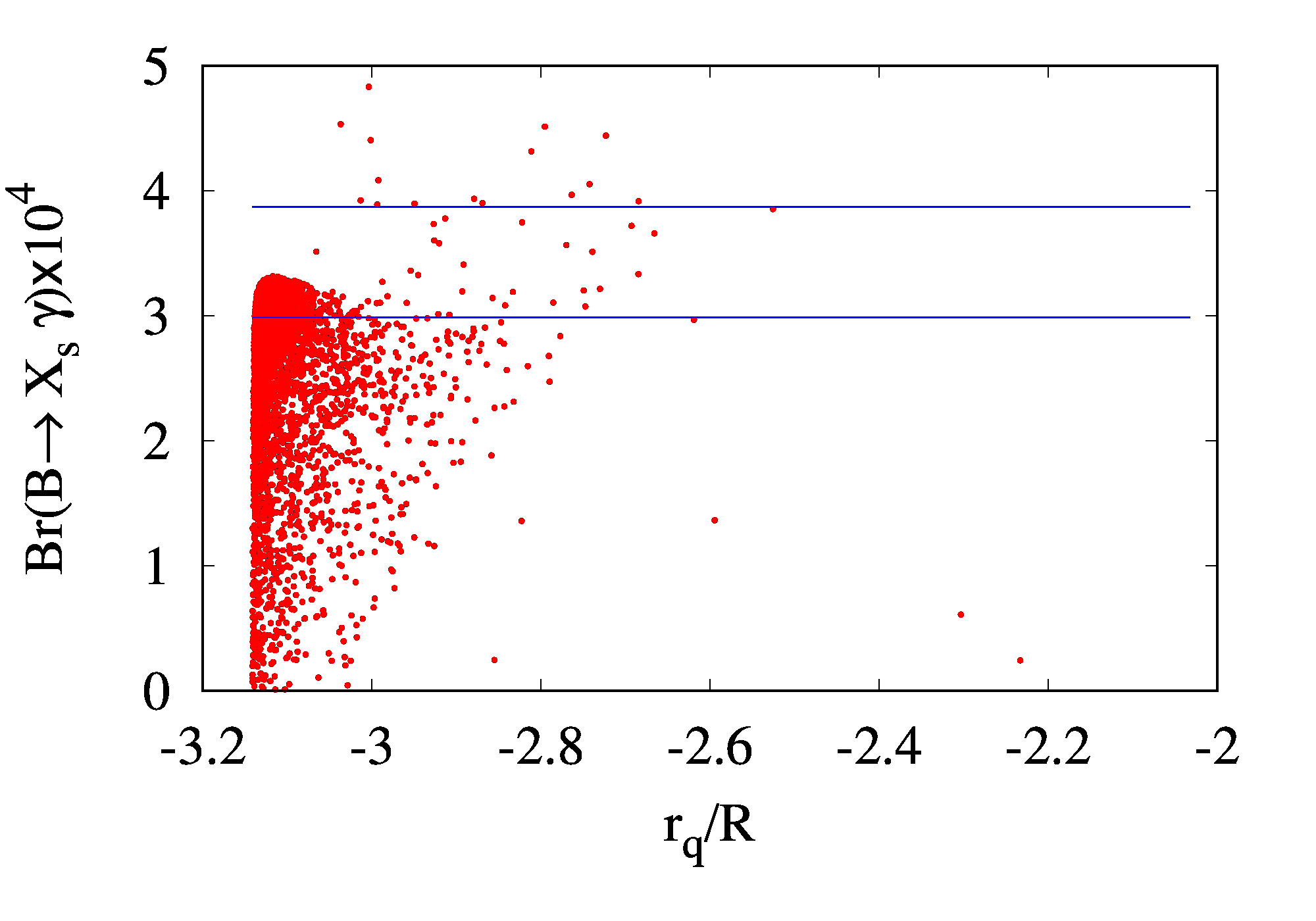} &
\includegraphics[scale=0.12]{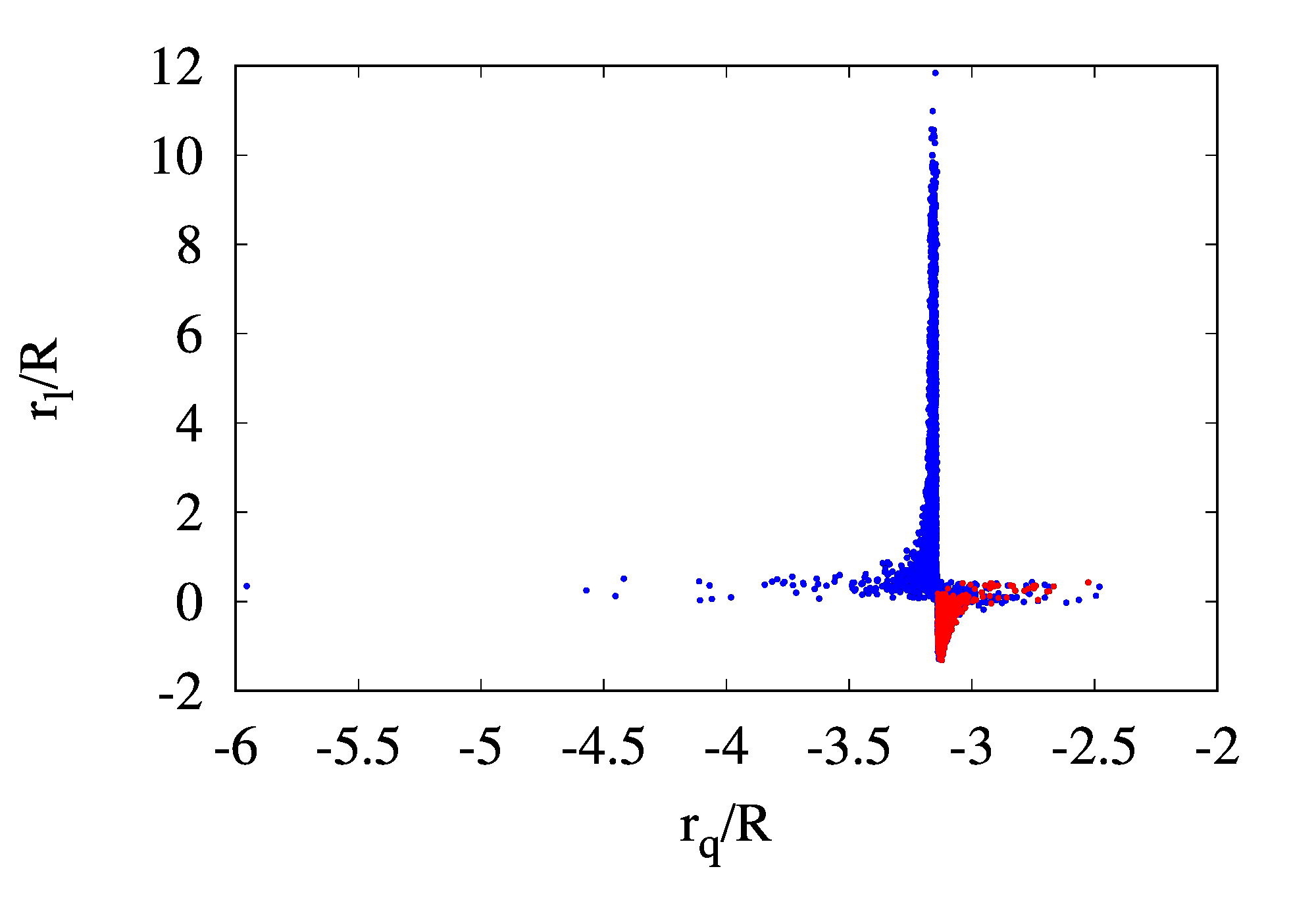} \\
(a) & (b) \\
\includegraphics[scale=0.12]{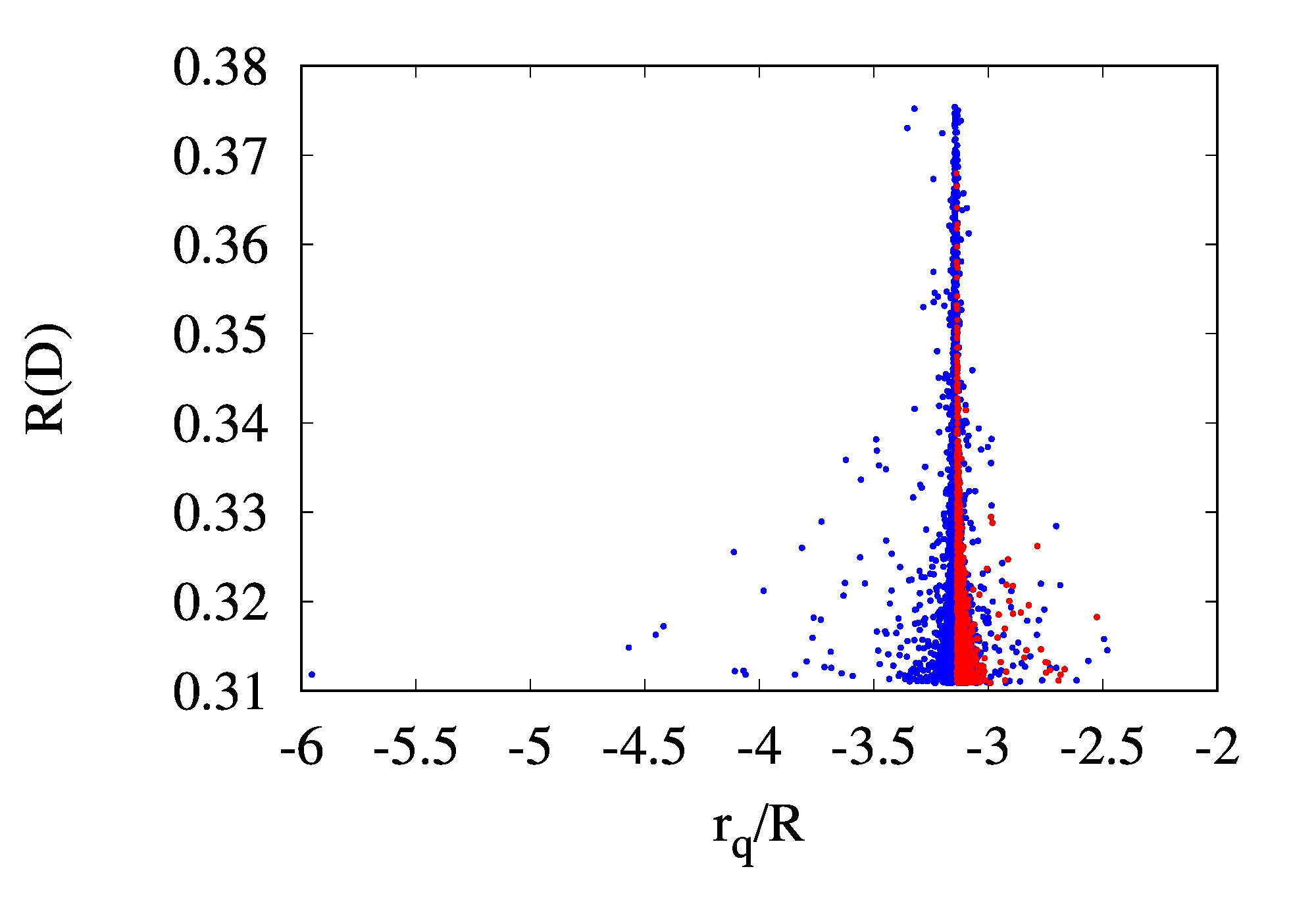} &
\includegraphics[scale=0.12]{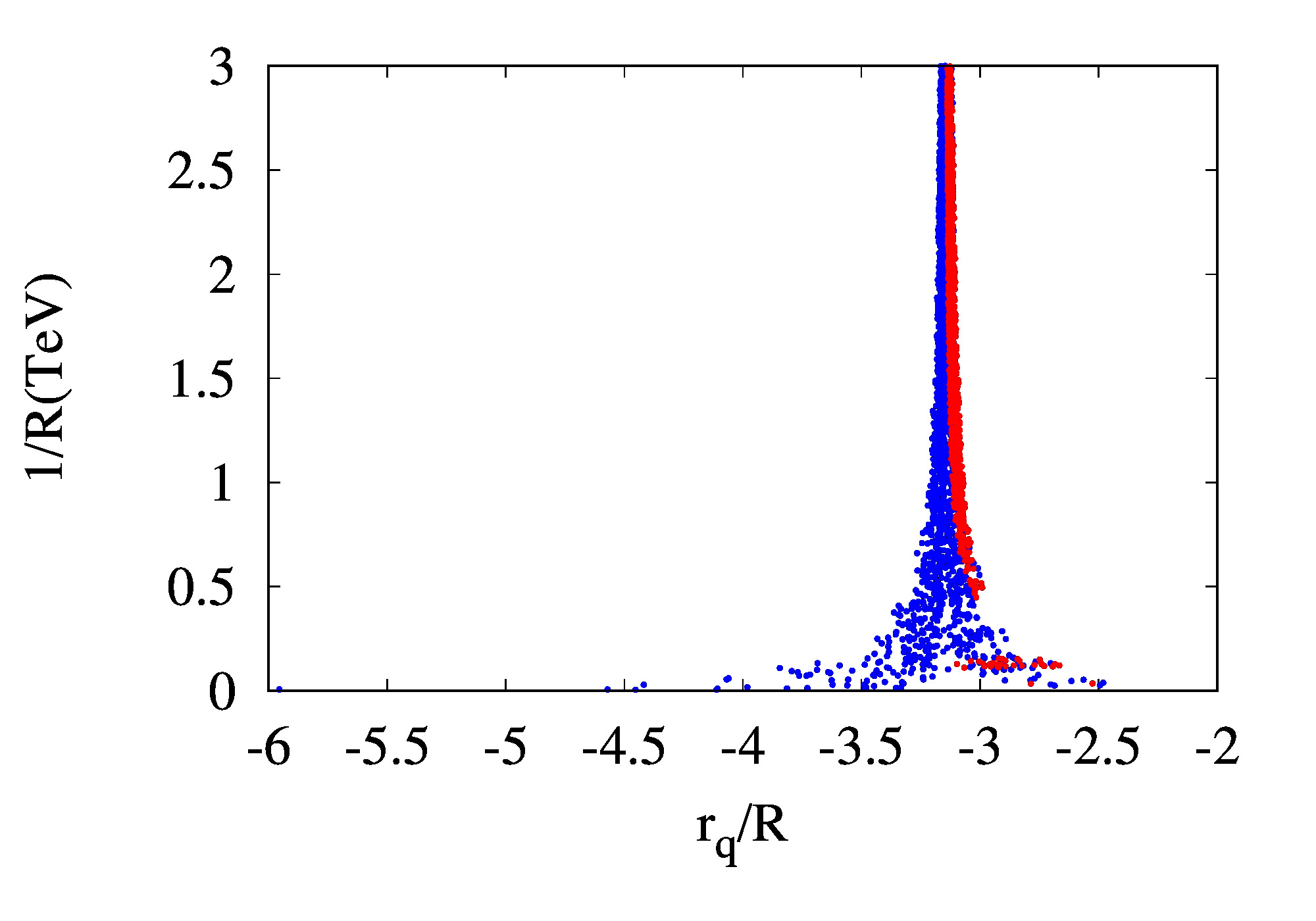} \\
(c) & (d) \\ 
\end{tabular}\caption{\label{Xsgamma}
Constraints from experimental results of ${\rm Br}(B\to X_s\gamma)=(3.43\pm0.21\pm0.07)\times 10^{-4}$
\cite{Datta17,HFAG14}. 
Panel (a) shows the values of ${\rm Br}(B\to X_s\gamma)$ with the points of Fig.\ \ref{F1} 
but with $-\pi<r_q/R$.
Blue horizontal lines are $2\sigma$ allowed bounds.
In panels (b), (c) and (d) allowed regions by ${\rm Br}(B\to X_s\gamma)$ (red) are given,
compared with those of Fig.\ \ref{F1} (blue). }
\end{figure}

In this analysis we do not consider explicitly possible constraints from the flavor changing neutral currents 
(FCNC) involving $b$ quark sector, but it needs some comments.
First, there is no FCNC at tree level because the BLT parameter $r_q$ is flavor independent.
The effective couplings of the even KK mode of gauge bosons and the SM quarks can be written as
the matrix in the flavor space \cite{Dasgupta}
\begin{equation}
G^{(n)}={\rm diag}\Big(g_{r_1}^{X^{(n)}}, g_{r_2}^{X^{(n)}}, g_{r_3}^{X^{(n)}} \Big)~,
\end{equation}
where $X^{(n)}=\gamma^{(n)}, Z^{(n)}$ and $r_1=r_2=r_3=r_q$ in our case.
The result is that $G^{(n)}$ is proportional to the identity matrix.
Second, $B_s\to\mu^+\mu^-$ and $B\to X_s\gamma$ are investigated in 
Refs.\ \cite{Datta16} and \cite{Datta17}, respectively.
The decay of $B_s\to\mu^+\mu^-$ or $B_s\to X_s\ell^+\ell^-$ involves both $I_n^q$ and $I_n^\ell$ 
while $B\to X_s\gamma$ does $I_n^q$ only.
In the former case since $I_n^q$ and $I_n^\ell$ both contributes to the process 
one can expect that the constraint on $I_n^\ell$ would not be as strong as that from the oblique parameters.
Actually in Ref.\ \cite{Datta16} the analysis was done with $r_q=r_\ell$. 
As one can see in Fig.\ 6 of \cite{Datta16}, the allowed parameter space for small $r_V$ is 
compatible with our results.
One point that must be noticed is that the lower limit of $R^{-1}$ is about a few hundred GeV,
which varies with $r_V$ and $r_{q,\ell}$ (see TABLE II of \cite{Datta16}).
In case of $r_q\ne r_\ell$, the allowed parameter space would be larger. 
In $B\to X_s\gamma$, only $I_n^q$ contributes to the process.
%
%
According to the Ref.\ \cite{Datta17}, dominant contribution comes not from $I_n^q$ but from other
overlap integrals, $I^n_{1,2}$ (see Eqs.\ (A10) and (A11) of Ref.\ \cite{Datta17}).  
The integrals contain a factor of $\sqrt{1+r_q/(\pi R)}$ and we restrict the range of $r_q$ as $-\pi<r_q/R$
in considering ${\rm Br}(B\to X_s\gamma)$.
In Fig.\ \ref{Xsgamma} we show the effects of ${\rm Br}(B\to X_s\gamma)$ on parameter space.
We only consider the lowest KK mode contributions for convenience.
As shown in Fig.\ \ref{Xsgamma} (b) positively large values of $r_\ell/R$ are not allowed.
We find that a considerable amount of parameter space is forbidden, 
but still the value of $\chi^2_{\rm min}$ is almost the same 
and the best-fit values of the observables remain also unchanged.
Third, $B_s$-$\Bbar_s$ mixing involves only $I_n^q$ and could provide a very strong constraint on $r_q$.
The SM prediction of the mass difference $\Delta M_s$ \cite{MsSM},
\begin{equation}
\Delta M_s^{\rm SM} = (20.01\pm1.25)/{\rm ps}~,
\end{equation}
is by $1.8\sigma$ larger than the measured value \cite{Amhis},
\begin{equation}
\Delta M_s^{\rm exp}=(17.757\pm0.021)/{\rm ps}~.
\end{equation} 
Usually the NP gives positive contribution and $\Delta M_s$ puts much stronger bounds on NP than before
because the updated SM prediction of $\Delta M_s$ gets larger \cite{Luzio}.
In our case $\Delta M_s$ is roughly $\sim (I_n^q)^4(m_W/M_{KK})^4$ where $M_{KK}$ is the mass
of the mediating KK particle.
At the $2\sigma$ level, 
$\Delta M_s^{\rm exp}/(\Delta M_s^{\rm SM}-2\delta\Delta M_s^{\rm SM}) -1\simeq 0.014$ where
$\delta\Delta M_s^{\rm SM}$ is the $1\sigma$ deviation of $\Delta M_s^{\rm SM}$.
One can naively guess that for $M_{KK}\sim 1~{\rm TeV}$ only order 1 of $I_n^q$ is allowed, 
which could severely constrain $r_q$.
Further study on this issue is necessary to scrutinize the model. 
%
\section{Conclusions}
In conclusion, we investigated the $B\to D^{(*)}$ anomalies in the nmUED model.
In the model, $n$th KK-modes of $W$-boson and scalar couple to a pair of zero-mode fermions
to result in nonzero NP Wilson coefficients.
We found that the nmUED model successfully fits the current data including $D^*$ polarizations, 
at the sacrifice of $r_\ell/R = r_q/R$.
The EWPT plays a significant role in the model.
Our main result is that the enhancement of the overlap integral in the quark sector 
is very crucial to explain the $B$ anomalies.
If there would be a quite strong constraint on the quark sector 
(e.g. from the neutral meson mixing or whatever)
then it could restrict the validity of the nmUED model seriously. 
We also found that the branching ratio ${\rm Br}(B_c\to\tau\nu)$ stays at a few percents, well below 10\%.
In our analysis $R(D)$ values have no overlap with the SM predictions at the $2\sigma$ level
while $R(D^*)$ touches the SM-allowed region.
Future measurements of more observables would check further the validity of the nmUED model. 


\end{document}